\documentclass[8.5pt,twoside,twocolumn]{article}
\usepackage[super,sort&compress,comma]{natbib} 
\usepackage{mhchem}
\usepackage{times,mathptmx}
\usepackage{sectsty}
\usepackage{balance} 

\usepackage{graphicx} 
\usepackage{lastpage}
\usepackage[format=plain,justification=justified,singlelinecheck=false,font=small,labelfont=bf,labelsep=space]{caption} 
\usepackage{epsfig}
\usepackage{epstopdf}
\usepackage{setspace}
\usepackage{oldgerm}
\usepackage{float}
\usepackage{amsmath}
\usepackage{amssymb}
\usepackage{subfig}
\usepackage[rightcaption]{sidecap}
\usepackage{tikz}

\graphicspath{{images/}}

\newcommand{\tauLJ}{{\tau_{\rm{LJ}}}}
\newcommand{\tauLB}{{\tau_{\rm{LB}}}}
\newcommand{\kB}{{k_{\rm B}}}
\newcommand{\Rg}{{R_{\rm g}}}

\begin{document}

\twocolumn[
  \begin{@twocolumnfalse}
\noindent\LARGE{\textbf{Chain deformation helps translocation}}
\vspace{0.6cm}

\noindent\large{\textbf{Farnoush Farahpour,$^{\ast}$\textit{$^{a}$} Azadeh Maleknejad,\textit{$^{b}$}  Fathollah Varnik,\textit{$^{c}$} and
Mohammad Reza Ejtehadi\textit{$^{a}$}}}\vspace{0.5cm}

\noindent \normalsize{Deformation of single stranded DNA in translocation process before reaching the pore is investigated. 
By solving the Laplace equation in a suitable coordinate system and with appropriate boundary conditions, an approximate solution for the electric field inside and outside of a narrow pore 
is obtained.
With an analysis based on ``electrohydrodynamic equivalence'' we determine the possibility of extension of a charged polymer due to the presence of an electric field gradient in the vicinity of the pore entrance.
With a multi-scale hybrid simulation (LB-MD), it is shown that an effective deformation before reaching the pore occurs which facilitates the process of finding the entrance for the end monomers. We also highlight the role of long range hydrodynamic interactions via comparison of the LB-MD results with those obtained using a Langevin thermostat instead of the LB solver.}
\vspace{0.5cm}
 \end{@twocolumnfalse}
  ]

\section{Introduction}

\footnotetext{\textit{$a$~Sharif University of Technology, Department of Physics, P.O. Box 11155-9161, Tehran, Iran. Fax: +98 21 66022711; Tel: +98 21 66164501; E-mail: farahpour@physics.sharif.ir}}
\footnotetext{\textit{${b}$~Institute for research in fundamental sciences (IPM), School of Physics, P.O.Box 19395-5531, Tehran, Iran. }}
\footnotetext{\textit{${c}$~ICAMS, Ruhr-Universitat Bochum, Stiepeler Strasse 129, 44801 Bochum, Germany. }}

\newcommand{\myepsilon}{\text{\usefont{OML}{cmr}{m}{n}\symbol{15}}}

The detailed dynamics and physics of polymer translocation through confined geometries, e.g. channels and pores, are not well-understood \cite{Meller2003,Nakane2002}.
In the last decades, experiments with single nanopores have been used to study translocation dynamics of single- and double-stranded DNA macromolecules through pores \cite{Meller2002}.
In these studies, a voltage bias drives the macroions through a single nanopore which can be a protein channel or a pore on an artificial solid-state substrate \cite{Meller2001,Chang2004,Storm2005}.
When a polymer traverses a pore in a membrane, it partially blocks the channel and so decreases the magnitude of the ion current passing through the pore due to an applied voltage.
By recording the ionic current modulation, one can thus monitor the polymer translocation indirectly \cite{Bezrukov1992,Meller2003,Meller2002,Chang2004}.

Due to the tiny size of the pore in comparison with the electrolyte chamber, the electrostatic voltage drops mainly inside the pore. Based on this observation, many studies neglect the effect of the electric field outside the pore on the dynamics of translocation \cite{Ambjornsson2002,Loebl2003,Chern2001}.
In such studies, it is also often assumed that polymer reaches the pore by a purely thermal diffusion process. 
Once sufficiently close to the pore, one of the end monomers then has to find the pore entrance. This marks the beginning of the translocation process during which the polymer is drawn into the pore by the electric force. However, the electric field in the vicinity of the pore entrance may have an important effect on DNA capture. In some studies in which the capture rate of translocation phenomenon is investigated, the biased diffusion of polyelectrolyte in a region close to the pore is considered \cite{Nakane2002,Meller2002,Wanunu2010,Grosberg2010,Muthukumar2010}.
In this paper, we show that this is not the only effect of the extended electric field outside the pore on translocation. Electric field is highly convergent on the cis side of the pore and divergent on the other side 
\cite{Nakane2002,Grosberg2010}. The fact that DNA is a highly charged flexible macromolecule, is a strong incentive 
to study the possibility of DNA deformation and polarization in a nonuniform electric field, which is prevalent close to the pore entrance. There is experimental evidence that, before translocation, DNA deformation and counterion decoupling occur in a capture region comparable to the hydrodynamic size of the 
chain \cite{Chen2004}. It is argued that, upon increasing the voltage, this mechanism can enhance the rate of diffusion-limited capture of DNA \cite{Wanunu2010}.
Deformation of DNA and its effect on the capture rate of the translocation process has been addressed in a number of previous theoretical works \cite{Wong2007,Grosberg2010,Muthukumar2010}.
To the best of our knowledge, however, this issue has not been tested in a direct simulation 
study so far.

Applying a mechanical \cite{Perkins1995,Smith1996}, electrical \cite{Bakajin1998,Han1999,Bertrand2007} or hydro-dynamical \cite{Brochard-Wyart1993,Ferree2003,Larson2006} force on a long polymer (in comparison to its persistence length) deforms the relaxed configuration of the polymer \cite{deGennes1999,DoiEdwards1986}.
According to this effect, a polyelectrolyte can be deformed from its coiled equilibrium configuration to an extended chain by a uniform (for an anchored polymer) or a non-uniform (for a 
freely moving polymer) electric field \cite{Randall2005,Kim2006}.
By simulating the DNA motion in a network of insulated cylinders, Tessier and coworkers have shown that DNA can deform electrophoretically in the entropic traps \cite{Tessier2002}.
Randall \textit{et al.} have also observed free DNA deformation in an electric field gradient.
These authors have shown that the electric field gradient created by insulator obstacles deforms DNA molecules.
Using Long's electrohydrodynamic equivalence \cite{Long1998}, they have also interpreted the observed dynamics of DNA deformation in an electric field gradient via a kinetic model \cite{Randall2005}.

Recently, Kowalczyk \textit{et al.} used an oblate spheroidal coordinate system to solve the Laplace equation inside the pore and have determined the electric field as a function of the pore shape parameters and the applied voltage \cite{Kowalczyk2011}. This assumption yields a reasonable description of access resistance of a pore before and after the insertion of a DNA molecule into the pore.
In this paper, we use this coordinate system and, after a simplification of the geometry of the pore, find an approximate expression for the electric potential in the entire space, just by setting the electrodes at infinity. We show that the corresponding electric field is consistent with recent experiments on the conductivity of nanopores \cite{Kowalczyk2011}.
We then determine the velocity gradient tensor and its eigenvectors and discuss the possibility of DNA deformation due to the presence of electric field gradient near the pore entrance.
Such deformation can help the ends of DNA molecules find the pore entrance and thus facilitate the capture process. Indeed, experiments show that DNA is more likely to be captured by the pore at the DNA's ends \cite{Storm2005-2}. A very recent study confirms these findings by a more accurate statistical analysis of the data \cite{Mihovilovic2012}.
 
Regarding the simulation methodology, we employ a hybrid lattice Boltzmann - molecular dynamics (LB-MD) technique. The coupling to LB is motivated by the need to account for the effect of long range ($\sim 1/r$) hydrodynamic interactions (HI).

For a single-stranded DNA (ssDNA) in dilute solutions which is studied in this paper, the Debye length,  $\lambda_{\rm D}$, and persistence length, $l_{\rm p}$ are of the same order of magnitude (for example, a typical value of $\lambda_{\rm D}$ in a $10~\rm{mM}$ solvent of monovalent salt is $3~\rm{nm}$ or larger while $l_{\rm p}\sim1\rm{nm}$\cite{Kesselheim2012, Kim2006}). So the local electric field disturbances by DNA and counterions movement can not be ignored \cite{Kesselheim2012}.
On the other hand, it seems that, even if we ignore this effect, we cannot neglect the importance of the arrangement of counterions around the DNA chain when studying the dynamics of a polyelectrolyte in a non-uniform electric field.
The ability of counterions to glide along the chain and detach and reattach to it, can give rise to a local polarization of the chain and thereby induce an extra interaction between the charged polyelectrolyte and the external field.
Therefore, we also study the effect of counterion rearrangement in a inhomogeneous electric field in the simulation part of this paper by applying a P3M method.

For simplicity, we neglect the effect of image charges and the related dielectric mismatch between the membrane and the solution. This is a good approximation if the Debye length is below the shortest characteristic distance between the DNA molecule and the membrane \cite{Kesselheim2012}. 

\section{Theory and Methods}
\subsection{Electric field}
To find an approximation for the electric field of a pore in the whole space, we use the oblate spheroidal coordinate system \cite{Kowalczyk2011} but with the electrodes at infinity.
We use $\mu$, $\nu$ and $\phi$ as the oblate spheroidal coordinates with a focal ring of radius 
$c$, where $\mu\in(-\infty,+\infty)$, $\nu\in[0,\pi]$ and $\phi\in[0,2\pi)$ (Fig.~\ref{2D-coordinate})). These selected coordinates map to the cylindrical ones by the simple relations: $\rho=c \cosh\mu \cos\nu$, $z=c \sinh\mu \sin\nu$ and $\phi=\phi$.
Surfaces of constant $\mu$ are confocal semi-ellipsoids, surfaces of constant $\nu$ are one-sheeted hyperboloids of revolution and surfaces of constant $\phi$ are semi-planes normal to the $x$-$y$ plane.
This set of coordinates map to the Kowalczyk's notation with $\mu=\sinh^{-1} \xi$, $\nu=\sin^{-1}\eta$; $\phi$ is the same in the two notations.
We assume that when the pore radius is greater than its length ($a>l$, where $a$ and $l$ are the pore radius and length, respectively), we can substitute a one-sheeted hyperboloid with $\nu=\nu_0$ for the membrane wall and the pore (Fig.~\ref{2D-coordinate}), where $\nu_0$ can assume any arbitrarily small value ($\nu_0\ll\pi/2$) which can be obtained from the pore radius, $a$, by the simple relation, 
$\nu_0=\cos^{-1}(a/c)$.
Since we want to study the deformation of a negatively charged polymer, it is preferable to consider the direction of electric field lines from the anode to the cathode (Fig.~\ref{DNA-pore}).
We adhere to this convention throughout this paper.
Assuming the cathode and anode sufficiently far from the pore center (negative and positive infinity, respectively), we have the following two boundary conditions:
\begin{figure}
\begin{center}
\begin{picture} (240,90)
\put(-10,-30) {
\subfloat {\label{2D-coordinate} \includegraphics[width=.5\columnwidth]{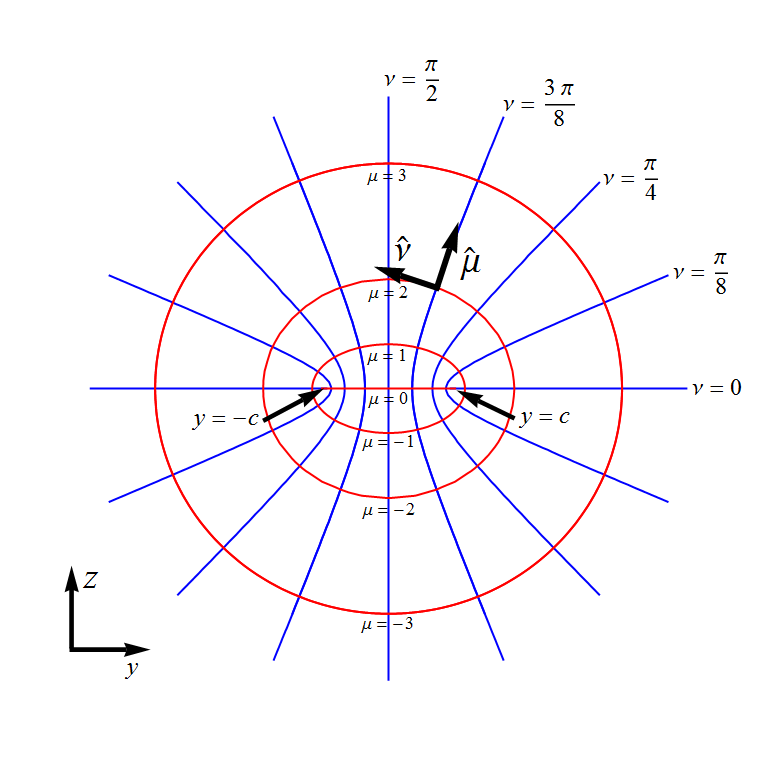}}
\subfloat {\label{DNA-pore} \includegraphics[width=.5\columnwidth]{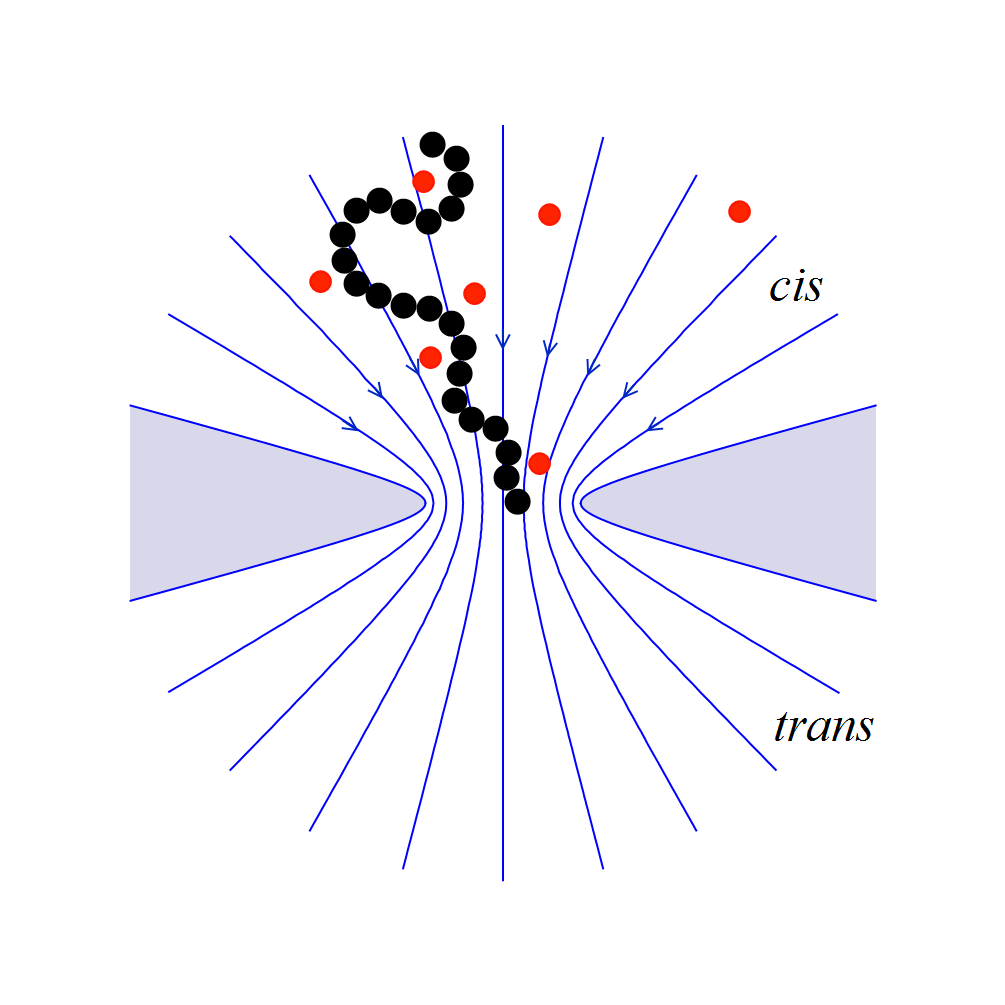}} }
\put(0,70){\textbf{(a)}}
\put(129,70){\textbf{(b)}}
\end{picture}
\end{center}
\caption{
a) One can mathematically simplify the geometry of the pore by considering it as a surface of a hyperboloid with $\nu=\nu_0$. b) The DNA translocation process is shown schematically, in the presence of an extended electric field (blue lines). DNA monomers are represented in black and some of its counterions in red.}
\end{figure}
\begin{equation}\label{boundary}
\begin{split}
\Phi(\mu,\nu,\phi)|_{\mu=\pm \infty} &=\mp \frac{V_{0} }{2}\\
\vec{E}(\mu,\nu,\phi).\hat{n}|_{\nu=\nu_0}&=\vec{\nabla }\Phi (\mu,\nu,\phi)|_{\nu=\nu_0}=0.
\end{split}
\end{equation}
Solving the Laplace equation with these boundary conditions, one obtains the following continuous electric potential in the entire space: 
\begin{equation}\label{potential}
\Phi (\mu,\nu,\phi)=\frac{V_{0} }{\pi } \tan ^{-1} (\sinh \mu).
\end{equation}
This potential is almost constant far away from the pore but varies rapidly close to it.
The equipotential surfaces are confocal semi-ellipsoids of constant $\mu$ with azimuthal symmetry.
The field lines lie on the hyperboloids, confocal with the hyperboloid representing the membrane wall (Fig.~\ref{DNA-pore}),
\begin{equation}\label{field}
\begin{split}
\vec{E}=-\frac{1}{h_{\mu} } \frac{\partial \Phi }{\partial \mu} \hat{\mu} =-\frac{V_{0} \hat{\mu}}{\pi a\cosh \mu\sqrt{\sinh ^{2} \mu+\sin ^{2} \nu} } , 
\end{split}
\end{equation}
where $h_{\mu}$ is the coordinate scale factor in $\hat{\mu}$ direction \cite{Kowalczyk2011}.

Although we have assumed $a>l$ to justify our simplification of the pore geometry, it is worthwhile to mention that one can construct pores with inverse aspect ratio ($l\geq a$) by adding two one-sheeted semi-hyperboloids (with $\nu_0\ll 1$) to the end of a cylinder of optional length ($l_{\rm c}$). Then, according to the uniqueness theorem, the solution of the Laplace equation is a continuous piecewise differentiable function comprised of three parts: first in the region $z>l_{\rm c}/2$, the electric field can be obtained from Eq.~(\ref{field}) after a translation by an amount of $l_c/2$ in $\hat{z}$ direction; for $-l_{\rm c}/2<z<l_{\rm c}/2$, the electric field is constant and homogeneous; finally for $z<-l_{\rm c}/2$, the electric field can again be obtained from Eq.~(\ref{field}).
We shall be mainly concerned with the chain dynamics before it reaches the pore. Thus, the details of the pore geometry are expected to have a minor effect on our results. We will return to a discussion of this aspect in the following sections.

Near the pore entrance, the functional form of the electric field is in good agreement with Nakane's solution \cite{Nakane2002}. For $r \gg a$, the electric field lines approximately lie on the lines asymptotic to the hyperboloids which are parallel to the $\hat{r}$ direction. Hence, far away from the pore, we reach the radial approximation for the electric field \cite{Wanunu2010}.

Using this extended electric field, we can calculate the conductance of the pore by computing the current obtained from the integration of the current density, $\vec{j}=\sigma_{\rm s} \vec{E}$, on any surface crossing the pore ($\sigma_{\rm s}$ is conductivity of the solution).
Equipotential surfaces with constant $\mu$ are good choices for this purpose. However, the special case of the $\mu=0$ surface turns out to be still more convenient. Using Eq.~(\ref{potential}) and after some simplification, one obtains:
\begin{equation}\label{current}
\begin{split}
J=\int \vec{j}.\hat{\mu} ds 
&= \sigma_{\rm s} \int _{\nu_0} ^{\pi/2} \int _{0} ^{2\pi} \frac{\partial \Phi}{\partial \mu} (\frac{h_\nu h_\phi}{h_\mu})d\nu d\phi \\
&=2 \sigma_{\rm s} V_0 a \frac{1-\sin\nu_0}{\cos\nu_0}.
\end{split}
\end{equation}
From this equation one can calculate the conductance of the semi-space from infinity to the entrance of the pore, $G_{\rm access}=4\sigma_{\rm s} a (1-\sin\nu_0)/(\cos\nu_0)$.
As discussed above, for the general case one can consider the pore as a combination of a 
cylinder and two semi-spaces with membrane walls coinciding with a hyperboloid.
Conductance of such a system is 
\begin{equation}\label{Conductance}
\begin{split}
G=(\frac{1}{G_{\rm channel}}+\frac{2}{G_{\rm access}})^{-1}=\sigma_{\rm s}a^2 
\big( \frac{l}{\pi}+\frac{a}{2}\frac{\cos\nu_0}{1-\sin\nu_0}\big)^{-1}.
\end{split}
\end{equation}
For $\nu_0\ll \pi/2$ this equation approaches the Hall approximation for the pore conductance \cite{Hall1975}, which provides an upper limit for the access resistance \cite{Kowalczyk2011}. It has been shown in previous studies that the asymptotic behavior of $G$ for the both limits of $a\ll l$ and $a\gg l$ is in good agreement with experimental measurements \cite{Kowalczyk2011}.

\subsection{Velocity gradient tensor}
\begin{figure}
\begin{center}
\begin{picture} (240,100)
\put(-10,-10) {
\subfloat {\label{fieldlines} \includegraphics[width=0.55\columnwidth]{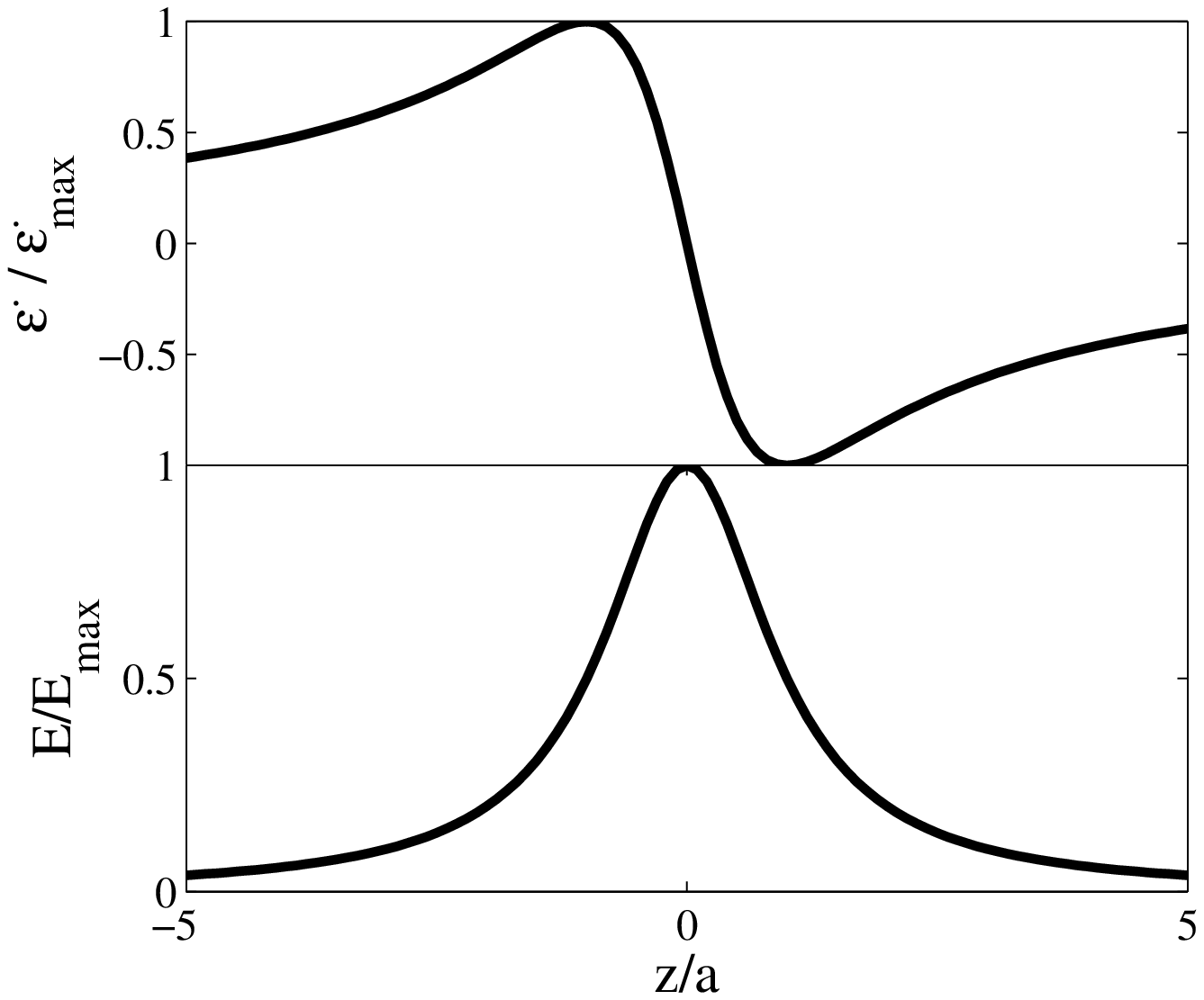}}
\subfloat {\label{elongation-axis} \includegraphics[width=0.45\columnwidth]{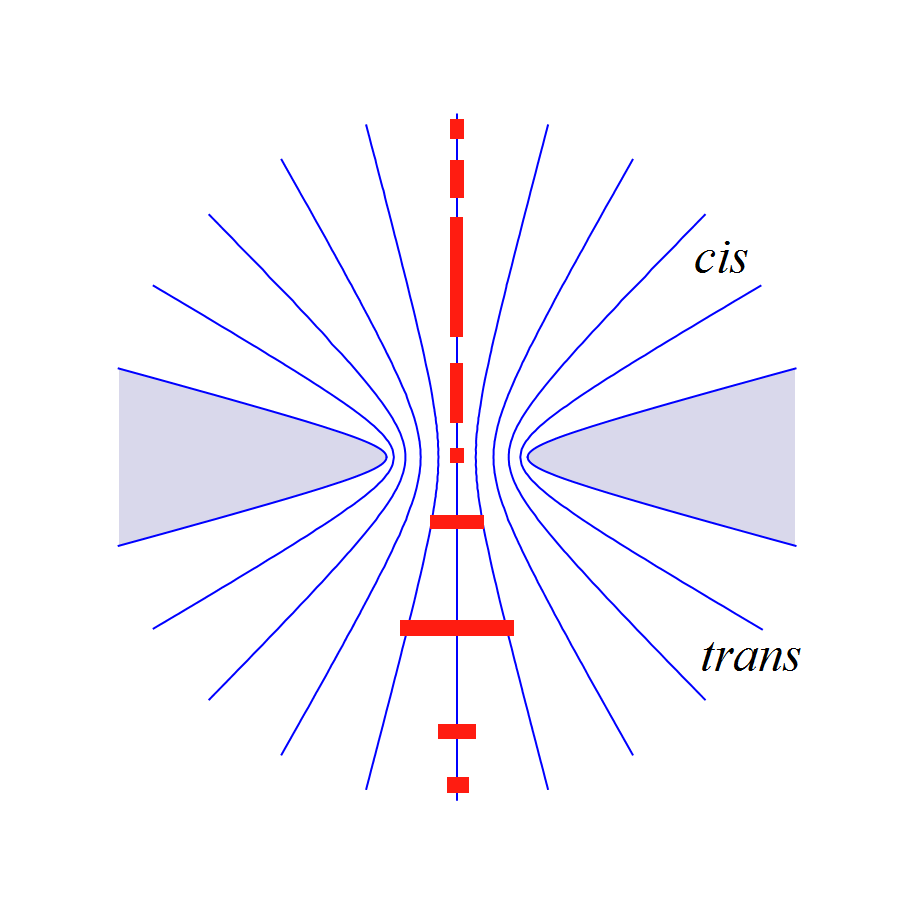}} }
\put(0,90){\textbf{(a)}}
\put(129,90){\textbf{(b)}}
\end{picture}

\end{center}
\caption{
a) Relative strain rate and relative electric field magnitude, normalized to their maximum values, versus relative distance to the pore, normalized to the pore radius (computed on the $z$ axis. See Eq.~(\ref{strain})). b) Direction and relative magnitude of extension axis 
along the centerline of the pore.
}
\end{figure}

To describe DNA deformation outside the pore we adopt the method introduced by Randall \cite{Randall2005}.
In our case, we generalize the method to three dimensions \cite{Deen1998}. Along this line, we first note that the symmetric rate-of-strain tensor governs the local deformation of fluid elements solely.
The local deformation in 3D extensional flows can be represented by a $3\times3$ symmetric matrix with orthogonal eigenvectors and real eigenvalues.
An ideal flexible object will exponentially extend/compress along the axis of extension/compression (with eigenvectors corresponding to positive/negative eigenvalues) at rates equal to the eigenvalues of the deformation tensor.
This ideal deformation is known as affine deformation \cite{Randall2005}.
The symmetric matrix of deformation is equal to the symmetric part of the velocity gradient tensor.
Electrophoretic velocity of a charged particle with mobility coefficient $\mu_{\rm el}$ in an electric field $\vec{E}(\mu,\nu,\phi)$ is equal to $\mu_{\rm el} \vec{E}(\mu,\nu,\phi)$.
Since $\nabla\times\vec{E}=0$, the gradient of this velocity vector field is symmetric, so its eigenvalues directly give us the strain-rate tensor.
The velocity  gradient tensor can be written as:
\begin{equation}\label{gradient_velocity_tensor}
 \mu_{\rm el} \vec{\nabla }\vec{E}=S\left(\begin{array}{ccc} {A} & {B} & {0} \\ {B} & {C} & {0} \\ {0} & {0} & {D} \end{array}\right), 
\end{equation}
where according to Eq.~(\ref{field}) we have,
\begin{equation}\label{gradient_velocity_tensor}
\begin{split}
&S=\frac{V \mu_{\rm el}}{a^{2} \pi } \frac{1}{(\sinh ^{2} \mu+\sin ^{2} \nu)^{2} \cosh ^{2} \mu},\\
&A= +\sinh \mu(\cosh 2\mu+\sin ^{2} \nu),\\ 
&B= +\cos \nu\sin \nu\cosh \mu,\\  
&C= -\sinh \mu\cosh ^{2} \mu,\\ 
&D= -\sinh \mu(\sinh ^{2} \mu+\sin ^{2} \nu). 
\end{split}
\end{equation}
Now we can obtain the electrophoretic strain rates $\dot{\varepsilon }(\vec{r})$ by calculating the eigenvalues
\begin{equation}\label{eigen_value}
\begin{split}
\lambda_{\pm} &=S\frac{A+C\pm\sqrt{(A-C)^{2} +4B^{2} } }{2},\\
\lambda_{\phi} &=SD,\\
\end{split}
\end{equation}
and the corresponding eigenvectors,
\begin{equation}\label{eigen_vectors}
\begin{split}
\vec{e}_{\pm} &=(\frac{\lambda_{\pm}}{S}-C)\hat{\mu}+B\hat{\nu},\\
\vec{e}_{\phi} &=\hat{\phi},\\
\end{split}
\end{equation}
of the tensor in Eq.~(\ref{gradient_velocity_tensor}). 
The eigenvectors specify the extension and compression axes of deformation.
$e_{\pm}$ are normal to $\hat{\phi}$ and lie in the $\mu$-$\nu$ plane.
For test purpose and without any loss of generality, we restrict ourselves to the 
pore axis, defined by $\nu=\pi/2$ ($z$ axis) \cite{Randall2005}.
On this axis, the eigenvalues are:
\begin{equation}\label{z_axis}
\begin{split}
&\lambda_{+}|_{\nu=\pi/2} = +2\frac{V\mu _{el} }{a^{2} \pi } \frac{\sinh \mu}{\cosh ^{2} \mu} (\vec{e}_{+}=\hat{\mu}),\\
&\lambda_{-} = \lambda_{\phi}|_{\nu=\pi/2} = -\frac{V\mu _{el} }{a^{2} \pi } \frac{\sinh \mu}{\cosh ^{2} \mu} (\vec{e}_{-}=\hat{\nu}, \vec{e}_{\phi}=\hat{\phi}).
\end{split}
\end{equation}
As seen from Eq.~(\ref{z_axis}), there is symmetry with respect to rotations around the pore axis. A similar symmetry is thus expected with regard to deformations in the plane perpendicular to the pore axis. On the other hand, a survey of the eigenvalues and the corresponding eigenvectors reveals that, in the cis chamber ($\sinh \mu>0$), DNA elongation occurs along the $\hat{\mu}$ direction while compression takes place in the normal plane.
To simplify the equations, we write the extension eigenvalue (strain rate along the pore axis) in Cartesian coordinates:
\begin{equation}\label{strain}
\dot{\varepsilon }(z)=+2\frac{V_{0} \mu _{el} }{a\pi } \frac{z}{a^{2} +z^{2} }.
\end{equation}
A polymer along the centerline of the pore experiences a pseudo-homogeneous extensional field \cite{Randall2005}.
Fig.~\ref{fieldlines} shows the strain rates (normalized to the maximum value) and the magnitude of the electric field along the $z$ axis.
It can be seen that, while the maximum value of the electric field occurs exactly at the middle of the pore, the strain rate (and so the gradient of the electric field) has a maximum at a somewhat larger distance, which depends on the pore shape.
The direction of the extension axis and relative elongation along the centerline of the pore are also shown schematically in Fig.~\ref{elongation-axis}.
The above discussion suggests that the linear alignment of a polyelectrolyte chain is enhanced by the electric field.

The integrated strain,
\begin{equation}\label{integrated_strain}
\varepsilon =\int \dot{\varepsilon } [\vec{x}(t)]dt,
\end{equation}
for a point-like molecule approaching the pore along the centerline can be calculated using Eq.~(\ref{strain}).
This enables us to compute $\varepsilon$ for a single chain using its center-of-mass trajectory, and draw the extension-strain curves and compare theoretical predictions with the results of simulations of the next section \cite{Randall2005}.

\subsection{Simulation}
The translocation of a ssDNA through a single nanopore, in the presence of an applied electric voltage, is simulated by employing the ESPResSo package \cite{Limbach2006}.
To take into account both the hydrodynamic and electrostatic interactions, we use a hybrid mesoscale method in which the lattice Boltzmann (LB) method \cite{McNamara1988,Higuera1989,Benzi1992,Qian1992,Gross2010,Gross2011} is applied to simulate the hydrodynamic effects in the system and a P3M algorithm is used to calculate full electrostatic interactions between charged monomers and counterions \cite{Deserno1998,Arnold2006}.
A bead-spring coarse-grained approach, in which groups of atoms are represented by a single interacting particle, is used for the MD part of the simulation. This enables a computationally efficient study of the translocation process. In experimental setups for the investigation of translocation, salt is always added to the system. Although most of our simulations have been performed in the absence of salt, we have also simulated a number of cases with added salt so that a conclusion can be drawn as to the effect of salt (see below).

The pore is represented by a solid substrate containing a circular orifice with symmetry axis along the $\hat{z}$ direction. The interactions between beads as well as between the chain and the membrane are written as the sum of five terms:
\begin{equation}\label{interaction}
U(\vec{r}^{N} )=U_{\text{FENE}} + U_{\text{WCA}} + U_{\text{wall}} + U_{\text{elec}} + U_{\rm{ext}},
\end{equation}
where $U_{\text{FENE}}$ and $U_{\text{WCA}}$ are respectively, the chemical bonds and excluded volume potential between the chain beads.
$U_{\text{wall}}$ is a hard-core interaction between the chain monomers and membrane (including the 
pore). 
The last two terms represent the electrostatic interactions and the
interactions with the external electric field. These are written 
individually as \cite{Lansac2004}:
\begin{equation}\label{interaction_terms}
\begin{split}
U_{\text{FENE}} &=\sum _{bonds}\frac{1}{2} k R^{2}\ln (1-(\frac{r_{ij}}{R})^2) ,\\
U_{\text{WCA}} &=\sum _{i}{\sum_{i<j}}^{\dagger}4\myepsilon([(\frac{\sigma_{ij}}{r_{ij} } )^{12} -  (\frac{\sigma_{ij}}{r_{ij} } )^{6} ] + \frac{1}{4}),\\
U_{\text{wall}} &={\sum _{iw}}^{\dagger}4\myepsilon([(\frac{\sigma_{iw}}{r_{iw} } )^{12} - (\frac{\sigma_{iw}}{r_{iw} } )^{6} ] + \frac{1}{4}),\\
U_{\text{elec}} &= \sum _{i,j}\frac{q_i q_j}{4 \pi \myepsilon_{\rm w} r_{ij}^2}\\
U_{\text{ext}} &= \sum _{i}q_i E(\vec{r}_i)
\end{split}
\end{equation}
where $r_{ij} =\left|\vec{r}_{i} -\vec{r}_{j} \right|$ is the distance between the sites $i$ and $j$ and $r_{iw}$ is the distance between the site $i$ and the wall.
$\sigma_{ij}=\frac{1}{2}(d_i+d_j)$ is the arithmetic mean of the radii of the beads involved in the interaction. 
The $\dagger$ superscript indicates that the summation is restricted to interparticle distances below the cutoff radius ($r_{ij}<r_{\rm c}$ and $r_{iw}<r_{\rm c}$).
We use the monomer diameter, $\sigma$, as the unit of length. The unit of energy is chosen to be $\myepsilon$. The time is given in units of $\tauLJ=\sigma\sqrt{m/\myepsilon}$, where $m$ denotes the mass of a monomer.
Unless explicitly stated, all the quantities in this paper are given in these reduced units (r.u.). 
Ions (counterions and salt ions) have a radius of $d_{\rm ion}=0.425\sigma$ \cite{Kesselheim2012}.
In the FENE potential, the bond stiffness and the maximum allowed bond length are set to $k=30$ and $R=1.5$, respectively. 
We use a cut-off distance of $r_{\rm c} =2^{1/6}\sigma_{ij}$ to consider only the repulsive part of the Lennard-Jones potential. Together with the FENE potential, this leads to a bond length of $r_{\rm{bond}}=1.2$. All simulations are performed at a temperature of $\kB T=1$ (the Boltzmann constant is set to unity ($\kB =1$)).
$\myepsilon_{\rm w}$ is the water permitivity. 
The external electric field, $E(\vec{r})$, is defined by Eq.~(\ref{field}).
In all the studied cases, we add an appropriate amount of monovalent counterions to the 
charged polyelectrolyte in order to neutralize the system.
To reach a desired salt concentration, we add a corresponding number of coions and counterions to the 
 system\cite{Kesselheim2012}.
A P3M method implemented in the ESPResSo package is used to calculate the electrostatic potential between charged beads. The screening effect of counterions is thus considered explicitly in the model.

An approximate map of the present model to a physical system leads to the real units of $\myepsilon=0.59~\rm{Kcal/mol}$, $\sigma=1.0~\rm{nm}$ and $\tauLJ=2\times10^{-11}~\rm{sec}$. 

Choosing $e$ (the elementary charge) as the electric charge unit, the real unit of voltage is $\sim25~\rm{mV}$.
In our simulations, depending on the situation, the applied external voltage varies between 8 to 12. Using the above unit conversion, this corresponds to 200--300 mV which lies nearly at the middle of the range in typical translocation experiments with solid-state nanopores.
Each monomer carries a reduced electric charge of $-3$ and each counterion and salt ion has a charge of $\pm1$. 
The Bjerrum length $l_{\rm B}=e^{2}/4 \pi \myepsilon_0 \myepsilon_{\rm w} k_{\rm B}T=0.70$ corresponds to the water Bjerrum length at room temperature \cite{Lansac2004,Grass2008}.

The size of the simulation cell is $L_x \times L_y \times L_z= 5 \Rg \times 5 \Rg \times 10 \Rg$, where $\Rg$ denotes the gyration radius of the chain. The pore has a radius of $1.5\sigma$ and an axial extension of  $5\sigma$. The pore radius is sufficiently large to allow the chain beads to pass through and small enough to prevent simultaneous passing of more than one bead (hairpin translocation).

Most of the simulations reported here include hydrodynamic interactions via a frictional coupling of the particle dynamics to the LB method. We use the D3Q19 model implemented in the package with a kinematic viscosity of $\nu=3.0$ and a fluid density of $\rho=0.864$. The LB grid spacing is $a_{\text{LB}}=1$ and the LB time step is $\tauLB = 0.05$. The coupling constant is set to $\Gamma_{\text{bare}}=20.0$ \cite{Grass2008,Kesselheim2012}. Again, all these quantities are given in the reduced (LJ) units.

Additionally, random fluctuations for particle and fluid act as a thermostat. We performed a number of benchmark simulations to make sure that this model with the mentioned parameters can reproduce the expected behavior of a single chain with HI. The results are in good agreement with those studies which have investigated the ability of MD-LB hybrid method in detail \cite{Ahlrichs1999,Ladd2009,Pham2009}.

\begin{figure}
\begin{center}
\begin{picture} (273,157)
\centering
\put(15,75) {\begin{tabular}{c}
\subfloat{\label{success-snapshot-1}\includegraphics[width=0.40\columnwidth]{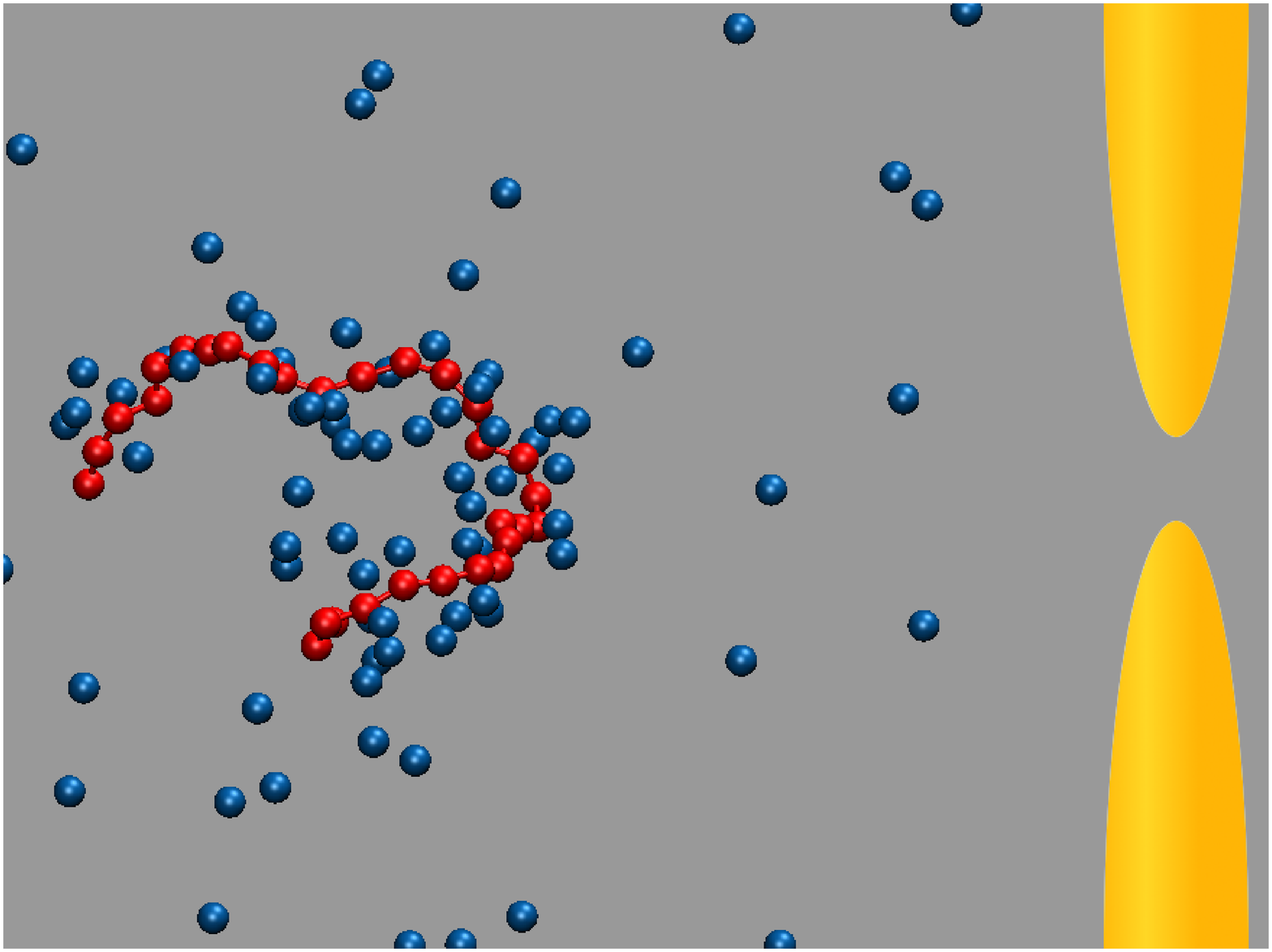}} 
\subfloat{\label{success-snapshot-2}\includegraphics[width=0.40\columnwidth]{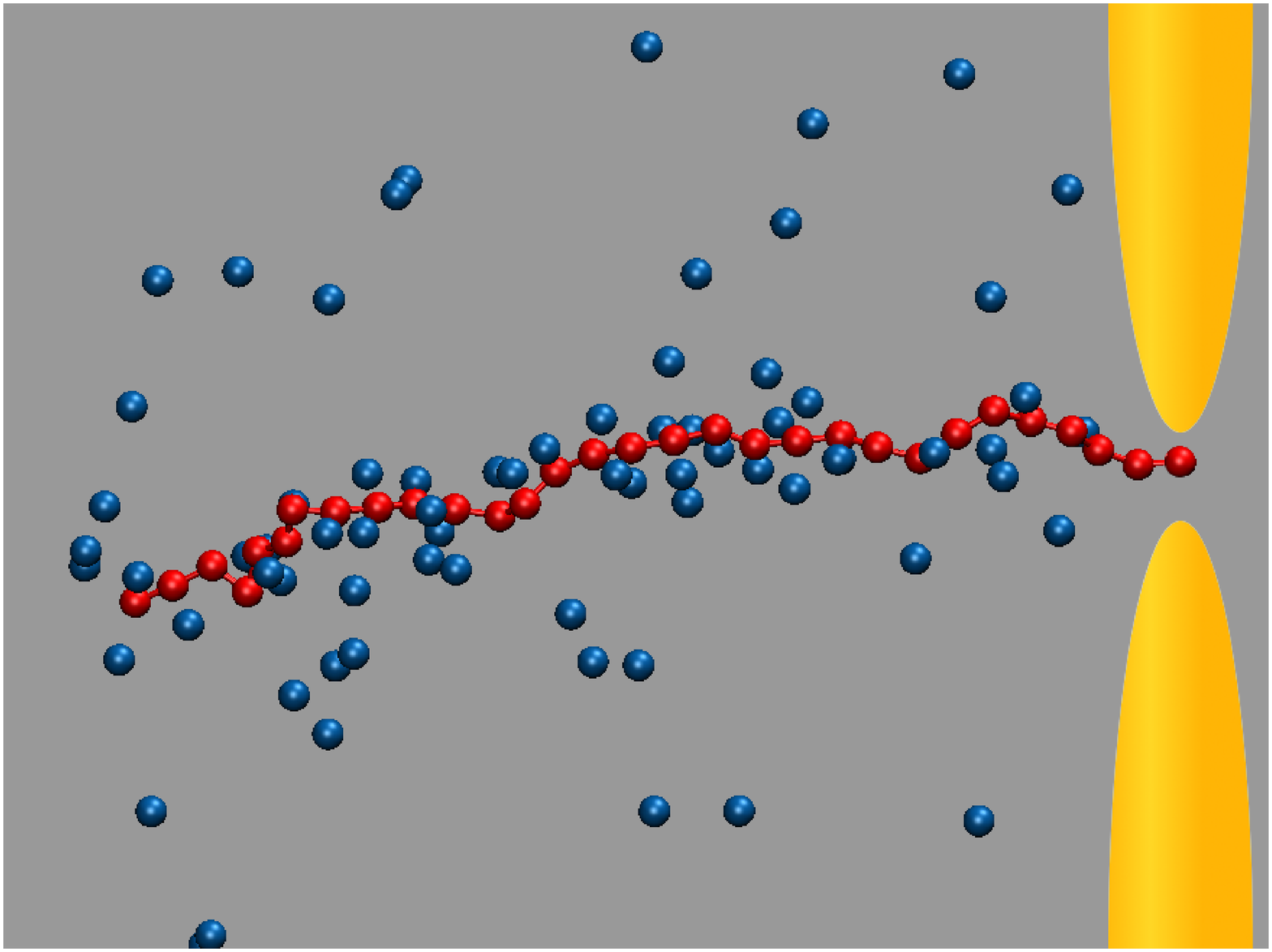}} 
\\
\subfloat{\label{unsuccess-snapshot-1}\includegraphics[width=0.40\columnwidth]{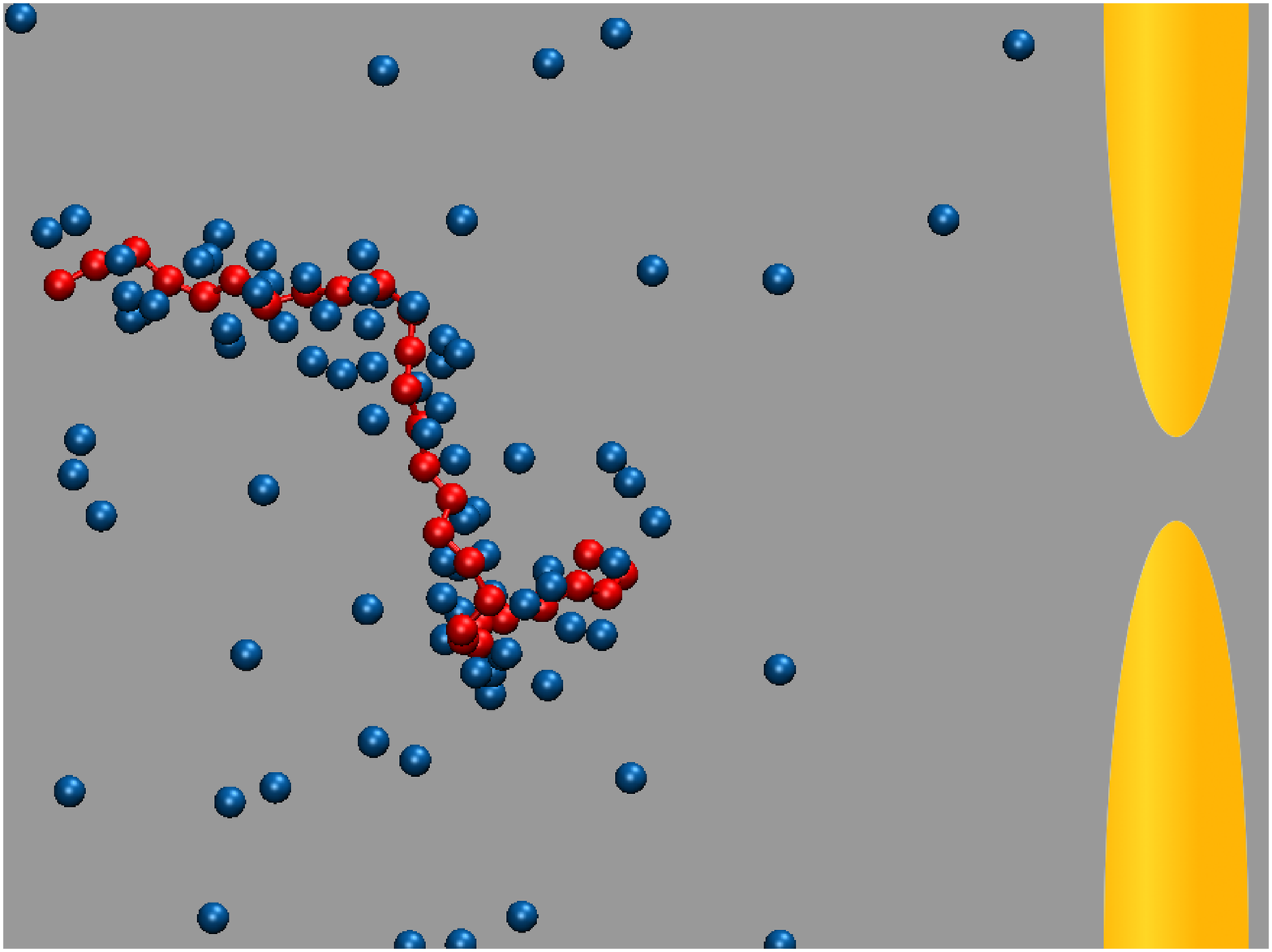}}
\subfloat{\label{unsuccess-snapshot-2}\includegraphics[width=0.40\columnwidth]{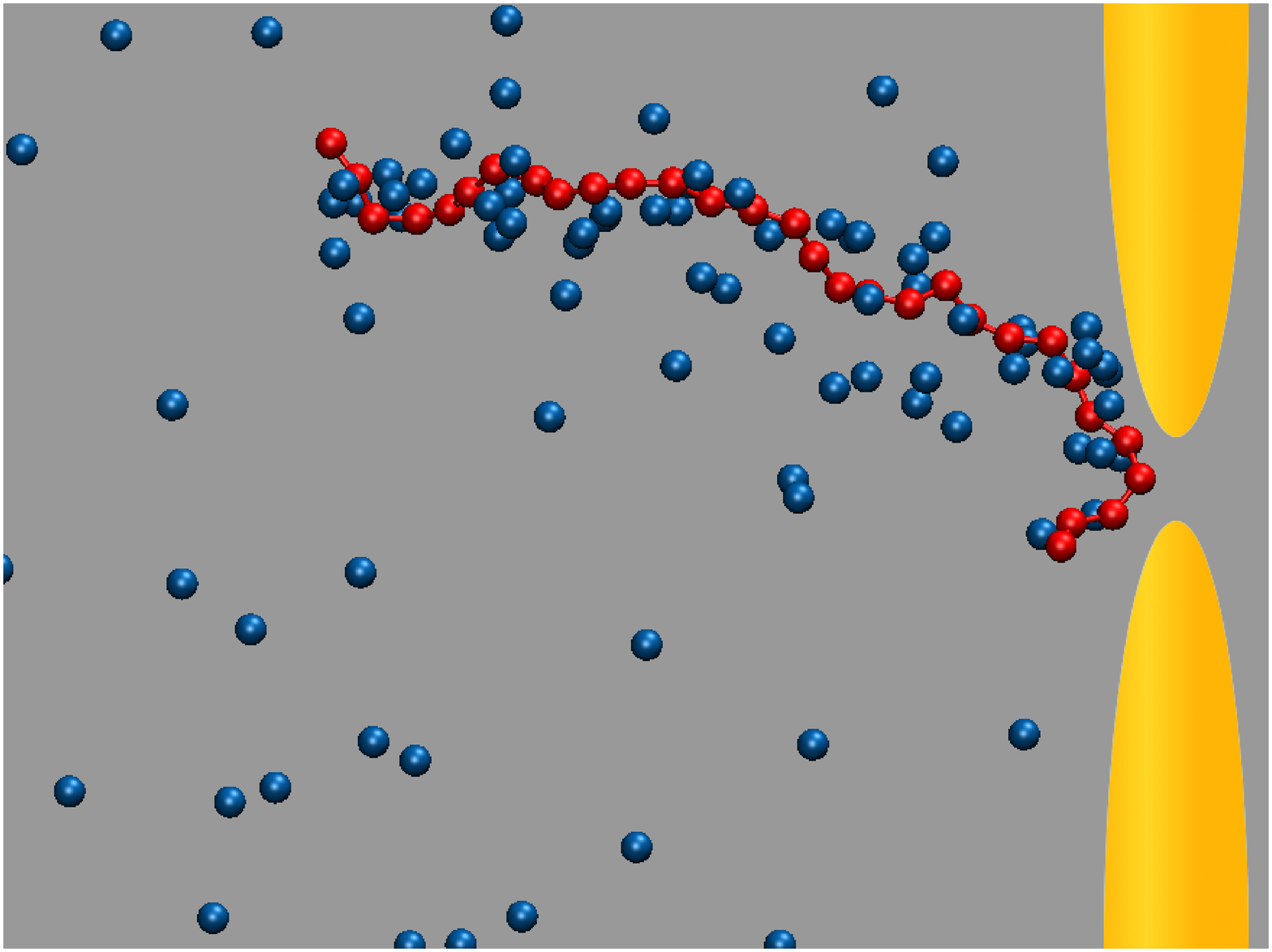}} 
\end{tabular}}
\put(25,143){\textbf{(a)}}
\put(124,143){\textbf{(b)}}
\put(25,55){\textbf{(c)}}
\put(124,55){\textbf{(d)}}
\end{picture}
\end{center}
\caption{
Typical configurations of a chain are shown in different situations. Upper panels (a and b) are selected from the snapshots of a successful translocation.
Extension along the electric field lines can be observed. Lower panels (c and d) show exactly the same situation for an unsuccessful try.
Length of the chain is 30 beads and the simulations have been done with a reduced voltage of $V^*=10$ and no added salt.
\label{snapshot}}
\end{figure}

The applied electric field is turned on after an initial equilibration run of duration $t=2\times10^6$ MD steps ($\Delta t=0.01$). This time is sufficiently large for the chain to reach its equilibrium configuration for all investigated chain lengths, $N$. The polymer's center of mass is initially located on the pore's axis at a distance equal to the capture radius, $r^*$, from the pore entrance. This is defined as the distance at which the electrophoretic interactions start to dominate and the polymer is attracted toward the pore. $r^*$ depends on the parameteres of the system such as the applied voltage, temperature, pore radius and the polymer length. It is obtained via a previously introduced method \citep{Grosberg2010} and is found to lie between a few up to several times $\Rg$ \citep{Grosberg2010}.

When DNA's center of mass reaches a distance greater than $r^*$, we transfer it radially toward the pore by applying a decaying radial force which acts on the monomers only for distances $r>r^*$ and is zero otherwise.

The DNA deformation, before entering the pore, is studied for some selected values of the electric voltage for chain lengths of $N=10,20,...,60$ in the absence of salt. For comparison, we have also performed a simulation for a $50~\rm{mM}$ salt solution at a chain length of $N=20$ monomer units.

\section{discussion}
Due to the electrophoretic forces acting on the charged monomers, the molecule is attracted toward the pore. At the same time, the chain is deformed as a result of the non-uniform nature of the electric field. Upper panels of Fig.~\ref{snapshot} show representative configurations of a chain before and during a successful translocation respectively. As can be seen, the polymer experiences a significant deformation before reaching the pore in such a way that one of the end monomers is captured by the electric field of the pore making the translocation possible.
\renewcommand{\arraystretch}{0.05}
\begin{SCfigure}
\begin{picture} (145,300)
\put(-6,148) {\begin{tabular}{c}
\cr {\includegraphics[width=4.9 cm]{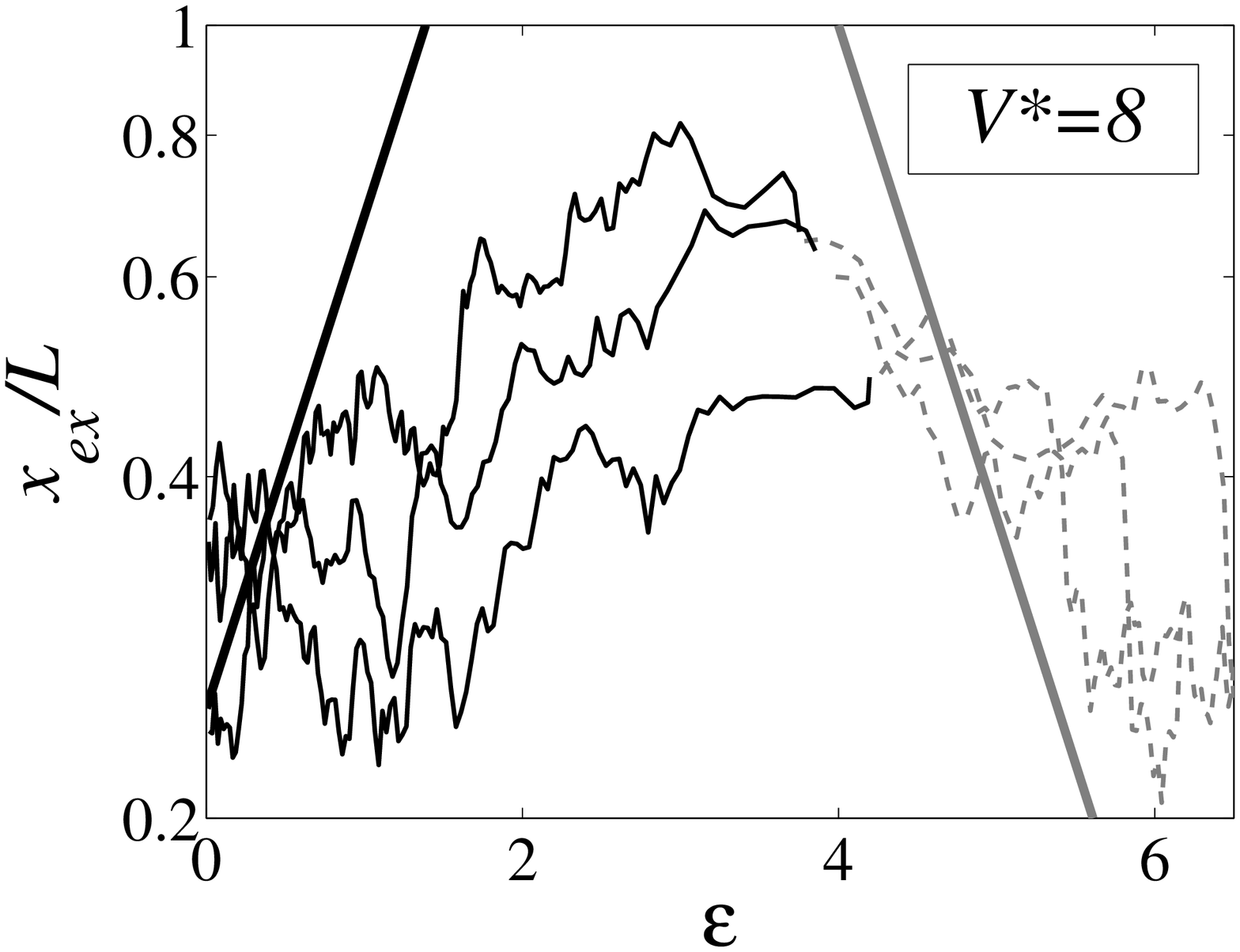}\label{extension_1}} \\
\cr {\includegraphics[width=4.9 cm]{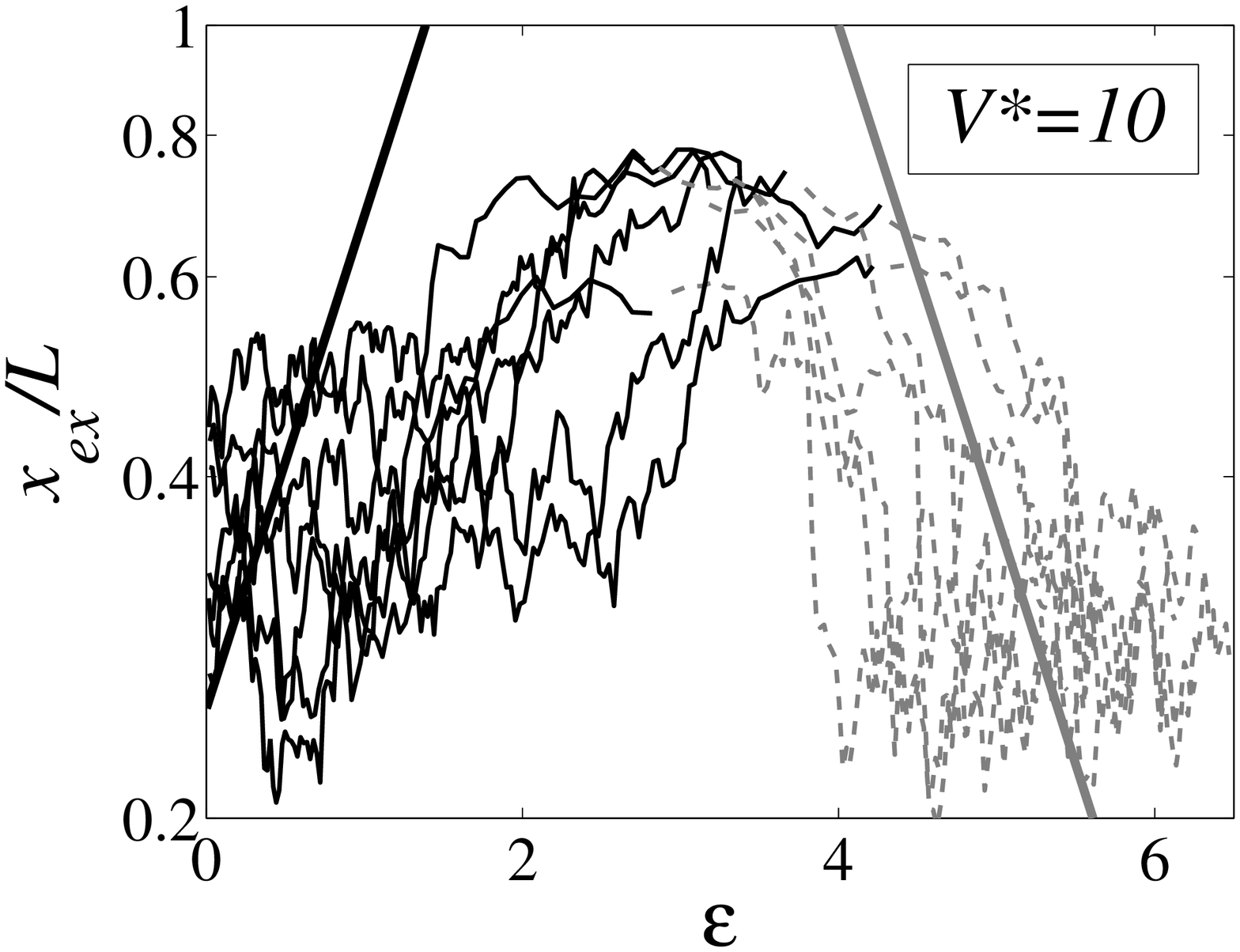}\label{extension_2}} \\ 
\cr {\includegraphics[width=4.9 cm]{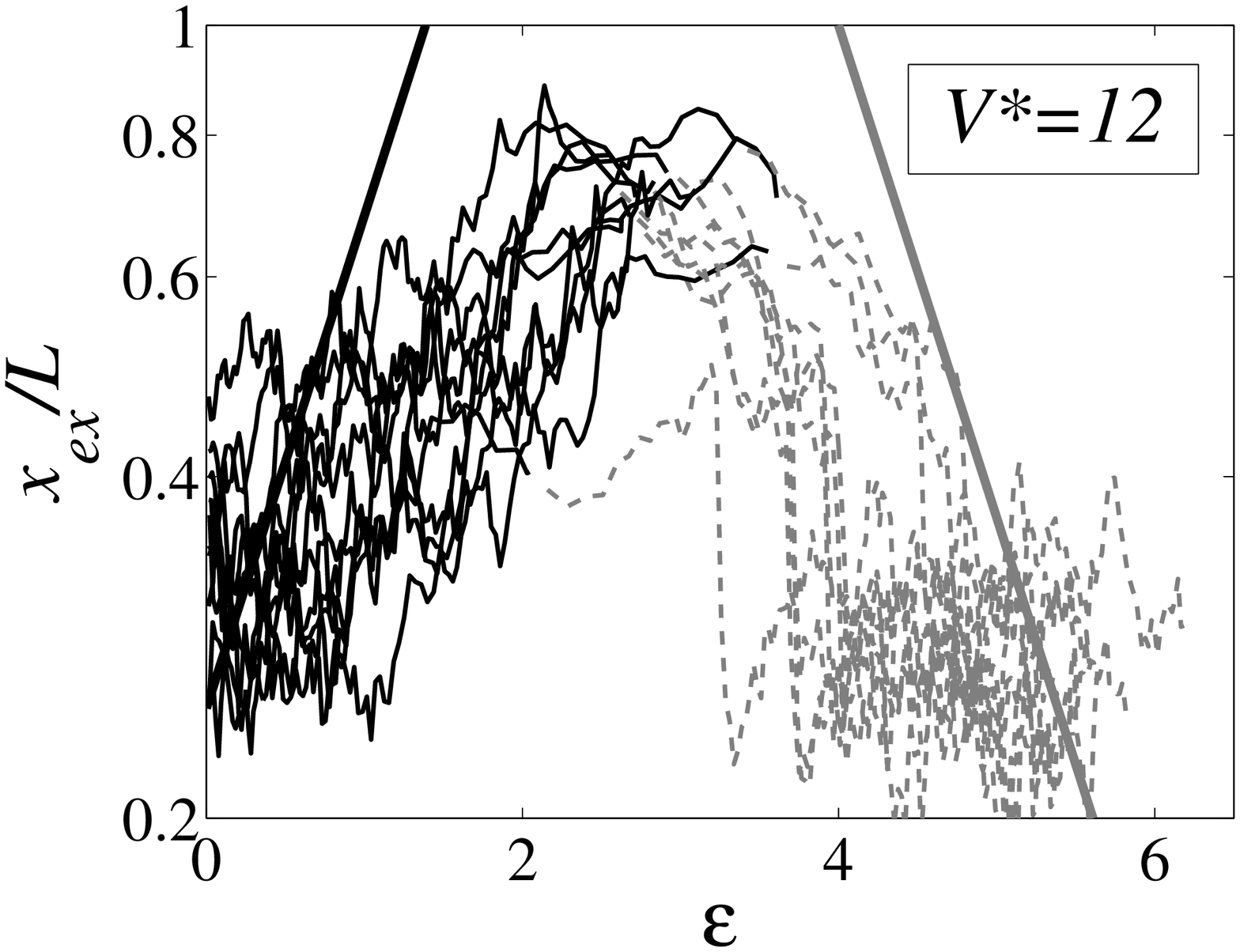}\label{extension_3}}  
\end{tabular} }
\put(26,287){\textbf{(a)}}
\put(26,187){\textbf{(b)}}
\put(26,87){\textbf{(c)}}
\end{picture}
\caption{
Extension-strain curves for different voltages during the translocation of DNA into the pore along the centerline of the pore. Each curve represents the trajectory of a single molecule.
Curves switch from solid black to dashed gray when the middle monomer of the chain crosses the midpoint of the pore.
Affine extension ($e^\varepsilon$, solid line) and compression ($e^{-\varepsilon}$, dashed line) scaling are shown for reference. In these figures, extension $x_{ex}$ is scaled to the polymer $L$ and strain is scaled to $\frac{V\mu_{el}}{4a\pi}\Delta t$.
The external applied voltage increases in successive figures:  a) $V^*=8$ b) $V^*=10$ and c) $V^*=12$. $N=20$ and there is no added salt.
\\
\label{extension}}
\end{SCfigure}

Even in an unsuccessful try (lower panels of Fig.~\ref{snapshot}), deformation occurs before reaching the pore and on average the chain elongates along the field lines. This increases the probability for an end monomer to be close to the pore entrance so that finding the entrance becomes easier for the chain in a subsequent try.

One can sketch the extension-strain curves by calculating the integrated strain (Eq.~(\ref{integrated_strain})) and the chain's longest extension in each time step as described in previous sections.
Relying on the assumption of affine deformation \cite{Randall2005}, we expect a linear relation between the logarithm of the chain extension and the integrated strain.
Such behavior can be observed in Fig.~\ref{extension}. In this figure, external voltage increases from $V^*=8$ to $V^*=12$. We have followed the successful translocations trajectory for those DNA molecules, whose path are nearly along the centerline of the pore.

As can be seen, the chain initially extends in the $z$ direction before reaching the pore.
On average, the maximum extensions occur just before translocation and coincide nearly with the 
maxima of extension-strain curves.
Just after the translocation where the sign of the eigenvalues of velocity gradient tensor changes, we expect a compression along the $z$ axis. 
This compression continues until the chain relaxes to its equilibrium configuration and then the strain rates and integrated strain magnitudes go to zero.
\begin{SCfigure}
\begin{picture} (145,230)
\put(-9,122) {
\begin{tabular}{c}
\subfloat{\label{Our_field}\includegraphics[width=5.3 cm]{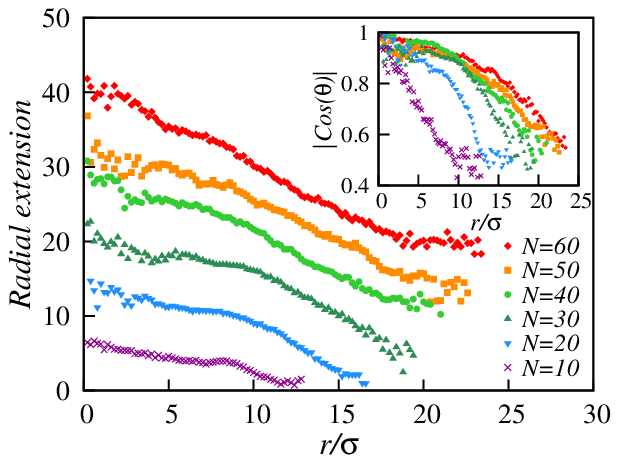}} \\
\subfloat{\label{Wanunu}\includegraphics[width=5.3 cm]{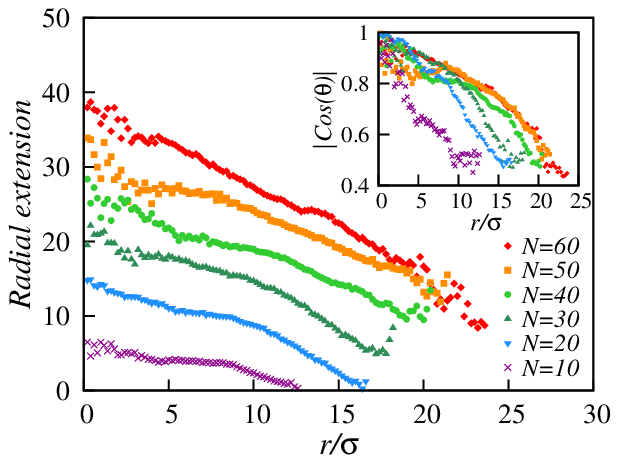}} \\ 
\end{tabular}
}
\put(22,223){\textbf{(a)}}
\put(22,97){\textbf{(b)}}
\end{picture}
\caption{
a) The radial extension of the chain versus radial distance to the pore (in units of $ \sigma$), for different lengths of the chain.
$V^*=10$ and there is no added salt. Inset shows $|\cos(\theta)|$ ($\theta$ is the angle between the extension axis and the electric field direction in the position of the polymer's center of mass) versus the radial distance.
Data are averaged over tiny windows of radial distance and over 100 simulations for each length.
b) The same as (a) but with the radial electric field applied outside the pore.
}
\label{enong}
\end{SCfigure}

To see the effect of this deformation on translocation phenomenon, we computed the eigenvectors and eigenvalues of the polymer gyration tensor during the motion of the polymer. The eigenvector corresponding to the largest eigenvalue of this tensor is the extension axis of the polymer. Radial extension is defined as the projection of the end-to-end distance vector on the radial vector which points from the pore to the center-of-mass of the chain.
In Fig.~\ref{enong}a, the radial extension of the chain (averaged over 100 independent simulations), is plotted versus distance to the pore. One can see from this plot that the extension of the chain increases as the chain approaches the pore.

To investigate whether this extension is along the electric field lines, we compute the angle, $\theta$, between the extension axis of the polymer and the electric field direction at the position of the polymer's center of mass. 
The inset in Fig.~\ref{enong}a shows $|\cos(\theta)|$ (averaged over 100 independent simulations) versus distance to the pore. One can see that the elongation along the field lines is nearly perfect in the vicinity of the pore for all chain lengths investigated.

Interestingly, we observe that the details of the emerging electric field only play a minor role for the behavior of the chain outside the pore. To illustrate this aspect, we show in Fig.~\ref{enong}b results on the chain deformation for a inhomogeneous but radial electric field \cite{Wanunu2010}. As a comparison of the panels (a) and (b) in Fig.~\ref{enong} reveals, the deformation behavior of the polymer is nearly the same for the both types of electric field investigates.

\begin{SCfigure}
\begin{picture} (150,95)
\put(-4,-10) {
\includegraphics[width=0.625\columnwidth]{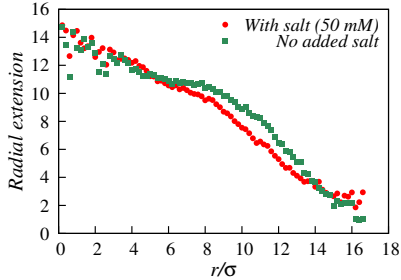}}
\end{picture}
\caption{
The radial extension of the chain versus the radial distance to the pore (in units of $\sigma$), with and without added salt. $N=20$ and $V^*=10$. 
}
\label{salt}
\end{SCfigure}

Regarding the effect of added salt, we investigated the elongation behavior of a chain in a solution of mono-valent salt with a concentration of $50 \rm{mM}$ ($N=20$) for the two situations with and without added salt. As shown in Fig.~\ref{salt},  the extension of the chain along the electric field does not seem to be strongly affected by the added salt. This may be due to the saturation of the ion cloud of the polymer with its counterions and the fact that our electric field is imposed externally with no contribution from the salt ions.

\begin{figure}
\begin{center}
\begin{picture} (240,175)
\put(-10,-8) {
\subfloat{\includegraphics[width=0.5\columnwidth]{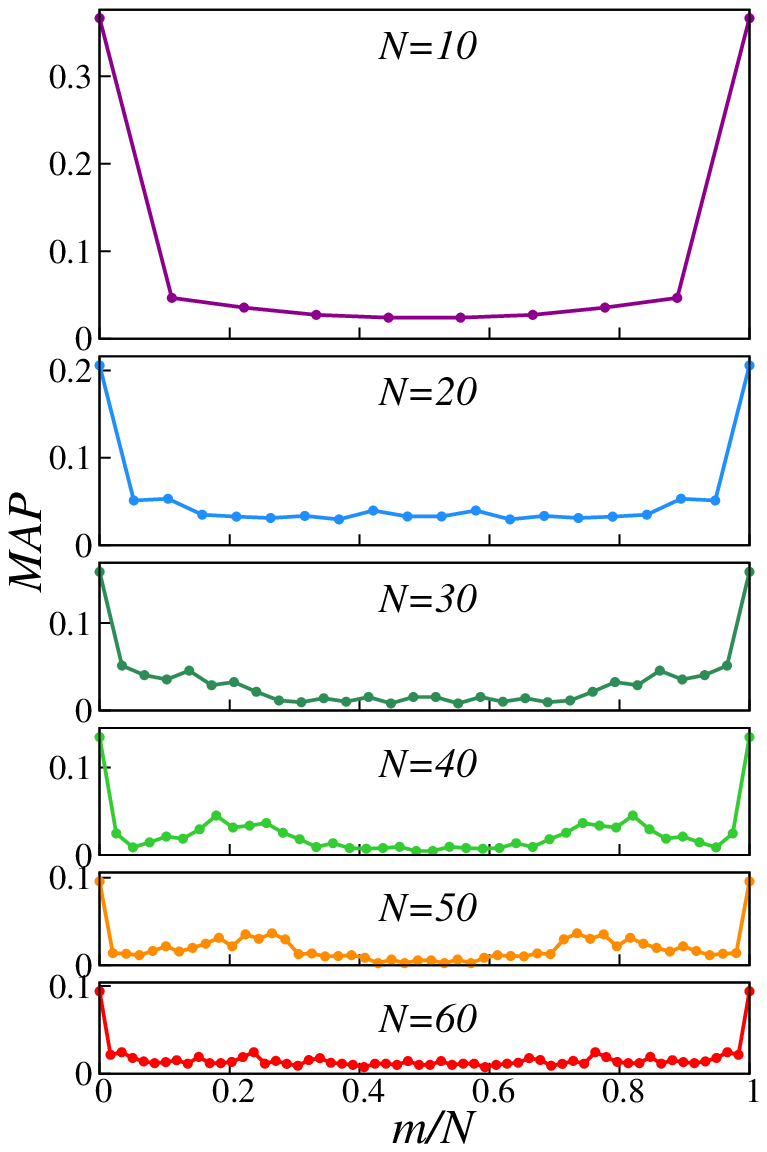}\label{first_reach_prob_WC}}
\subfloat{\includegraphics[width=0.5\columnwidth]{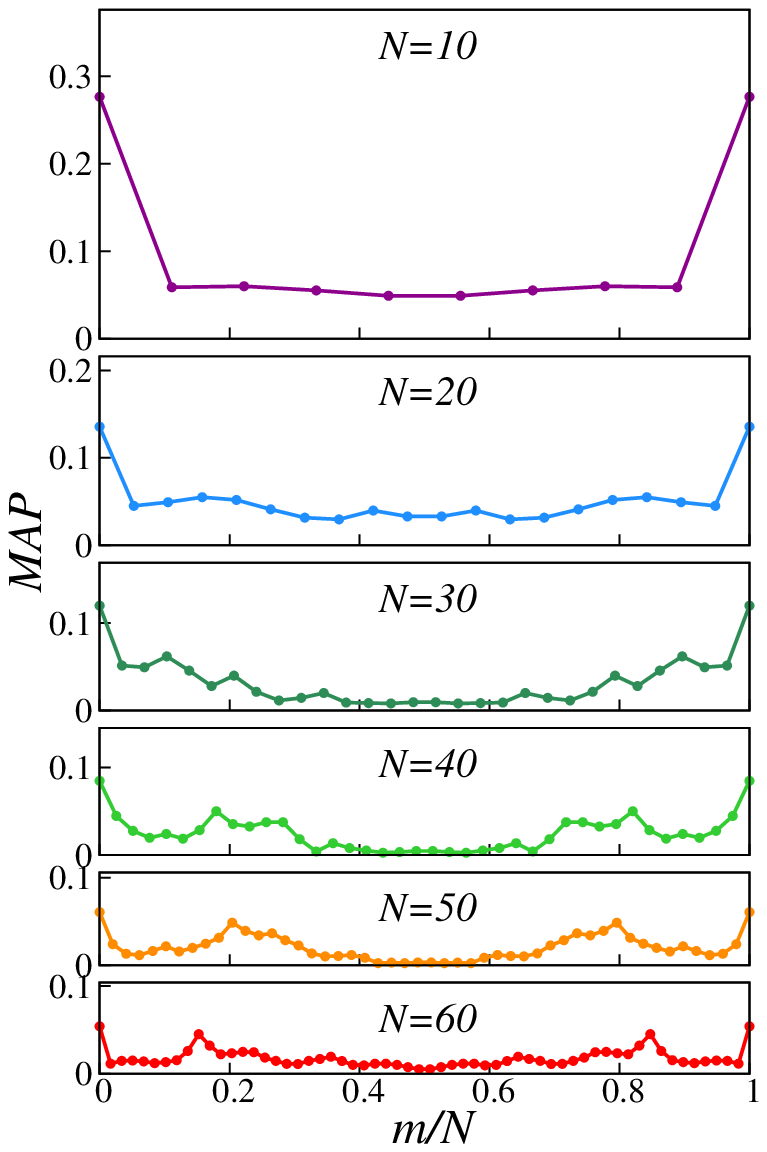}\label{first_reach_prob_NC}} }
\put(0,-10){\textbf{(a)}}
\put(124,-10){\textbf{(b)}}
\end{picture}
\end{center}
\caption{
a) Monomer approaching probability (MAP) along the chain for different lengths with explicit counterions for a voltage of $V^*=10$ and no added salt.
For better comparison between different lengths, we normalized the monomer ID (m) to the degree of polymerization (N).
b) Same diagrams as in a) but using mean-field effective charges instead of explicit counterions.
}
\end{figure}

To see the effect of the extended electric field and the related chain deformation on the capture process, we define the monomer approaching probability, MAP$(m)$, of monomer $m$. MAP$(m)$ gives the number of cases (divided by the total number of polymer approaches) in which the monomer $m$ is the first monomer of the chain which reaches a distance of the pore smaller than $2.5\sigma$, starting from an equilibrated configuration. 
The obtained MAP is plotted in Fig.~\ref{first_reach_prob_WC} versus the normalized monomer ID, $m/N$. The plot clearly shows that the end monomers have a higher probability to reach the pore entrance as compared to other monomers of the chain. We remark that this effect is significantly more pronounced than the well-known wall-induced enrichment of the end monomers in the vicinity of a planar wall \cite{Varnik2000b}. We thus attribute the main contribution to this behavior to the deformation of the chain in the extended electric field. 
One can also see that the relative behavior of the MAP is nearly the same for all the chain lengths investigated. When we compare this result with the rather smooth variation of the distribution of the end monomer with distance from the polymer's center of mass in its relaxed configuration \cite{Witten1981}, we find a significant deviation which mainly arises from the interaction of the electric field and the charged polyelectrolyte. The general behavior of the MAP seen in Fig.~\ref{first_reach_prob_WC} is consistent with experimental reports on the enhanced capturing probability of end monomers \cite{Storm2005-2,Mihovilovic2012}.
\begin{figure}
\begin{center}
\includegraphics[width=0.8\columnwidth]{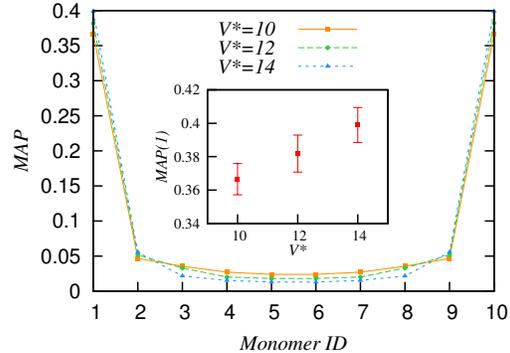}
\end{center}
\caption{\label{MAP_tot}
Monomer approaching probability (MAP) for a chain length of $N=10$ for three selected values of the voltage. The inset shows in detail the variation of MAP for the end monomers. Note the ascending trend toward greater MAP with increasing voltage. Error bars give the standard deviation of independent simulations.}
\end{figure}

Next we investigate the effect of counterion distribution. For this purpose, we remove the explicit counterions from the simulation but attribute a constant effective charge (which is length-dependent in the investigated $N$-range) to all the monomers. This effective charge is estimated from equilibration simulations with explicit ions, using the method introduced in a previous study on polyelectrolytes\cite{Grass2008}.

In Fig.~\ref{first_reach_prob_NC}, the monomer approaching probability is plotted versus the normalized monomer ID. One can see that eliminating the freedom of counterions to move and rearrange in the inhomogeneous electric field decreases the MAP of the end monomers especially for longer chains.
One possible explanation for this observation is that the monomer-counterion bonding strength depends on the position of the monomer along the chain's backbone. While the counterions close to middle monomers are more confined and thus cannot detach so easily from the chain, it is relatively easy for the counterions close to the end monomers to leave the chain. This increases the effective charge of the end monomers, thereby enhancing the alignment of the chain along the electric field lines. This interpretation is in line with a previous report on the distribution of the condensed monovalent counterions on the chain \cite{Wu2011}.

As seen in Fig.~\ref{MAP_tot} and the corresponding inset, increasing the voltage enhances the MAP for the end monomers, while it decreases the MAP for the inner monomers. Again, this is in line with the interpretation that the electric field tends to stretch the polymer. The larger the applied voltage, the larger this stretching and alignment effect, which increases the MAP of the end monomers but decreases that of the monomers further away from the both ends. This is also in agreement with those studies which report a bias for unfolded translocations that increases with the applied voltage \cite{Chen2004,Mihovilovic2012}.

\begin{figure}
\begin{center}
\includegraphics[width=0.8\columnwidth]{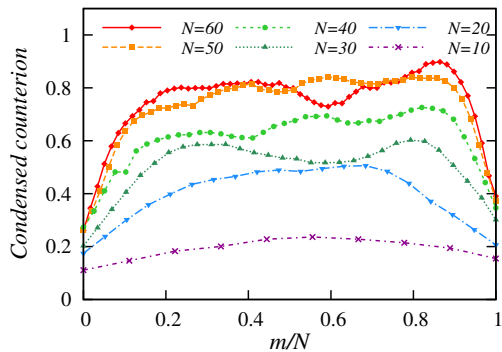}
\end{center}
\caption{\label{counterion}
Distribution of condensed counterions along the chain's backbone for different chain lengths, $N$, for the case of $V^* = 10$ and no added salt. The monomer ID ($m$) is normalized by the chain length.}
\end{figure}

In order to test this idea, we have counted the number of condensed counterions along the chain, when it is aligned with the electric field lines. We assign the ID number 0 to the end monomer which is closer to the pore mouth.
As it is visible from Fig.~\ref{counterion}, the symmetry in the number of condensed counterions along the chain is broken and there are more condensed counterions around the monomers away from the pore.

\begin{figure}
\begin{center}
\includegraphics[width=0.9\columnwidth]{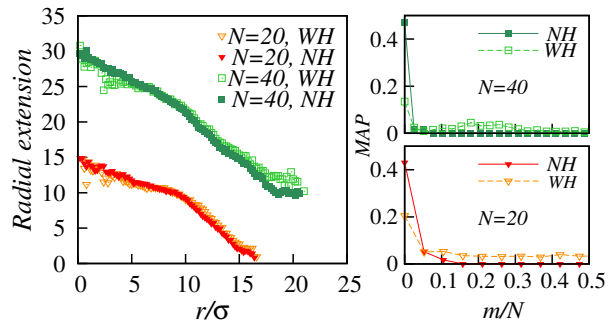}
\end{center}
\caption{\label{NH}
Effect of hydrodynamic interactions on polymer deformation. The radial extension of the chain is depicted versus radial distance from the pore for two different chain lengths of $N=20$ and $40$. The labels WH (NH) stand for simulation results obtained with (without) HI. $V^*=10$ and there is no added salt.
The right panels show the MAP for one half of the chain.
}
\label{HI-effect-on-extension}
\end{figure}

In addition to the above discussed effects, long range hydrodynamic interactions can give rise to collective motion and thereby influence the translocation process. Focusing on this aspect, we have also performed a series of simulations without HI so that a comparison with the results presented so far may allow to identify the effects arising from HI. In these new set of simulations, we substitute the lattice Boltzmann solver by a Langevin thermostat and thus eliminate the long-range hydrodynamic interactions. We set the friction parameter of the thermostat to $\Gamma_0 = 11.8$  in order to match the single-particle mobility of the Langevin system to the one with full HI. Figure \ref{HI-effect-on-extension} compares the results obtained from these simulations to those including hydrodynamic interactions. As seen from this figure, the effect of hydrodynamic interactions on chain deformation is quite small. However, a survey of the monomer approaching probability (right panels of Fig.~\ref{HI-effect-on-extension}) shows that neglecting HI leads to a significant increase of the MAP  of end monomers, reaching values as high as $0.48$. The fact that such a high MAP for end monomers is not in agreement with experiments \cite{Mihovilovic2012} shows that HI are quite important for a reliable study of translocation dynamics, even though its effects on chain deformation may be quite hard to resolve.

So far we have only considered inhomogeneous electric fields. It has, however, been reported that a polyelectrolyte with counterions in a homogeneous electric field greater than a critical value, $E_{\rm c}$, deforms and elongates in the direction of the field \cite{Netz2003,Frank2008,Wu2011}. Now, sufficiently close to the pore, the inhomogeneous electric field may become larger than $E_{\rm c}$ so that, at least in principle, the field strength itself, ignoring its inhomogeneous nature, can be the main reason for the observed deformation. It is, therefore, interesting to study the effects on the translocation process which arise from the inhomogeneous character of the electric field.

Since the previous set-up is not suitable to study the effect of all the above factors (electric field and its gradient, counterion rearrangement and HI), we devise a simulation setup which allows us to study the effects of both a uniform and a non-uniform field. For this purpose, we build a periodic box of size $L_x \times L_y \times Lz = 103 \times 69 \times 69$ without a pore and put a polymer of length $N=30$ in the middle of the box. After $10^6$ equilibrium steps, a homogeneous electric field of $\vec{E}=-E_0\hat{x}$ is switched on and the polyelectrolyte starts to move in the positive $\hat{x}$ direction. This homogeneous field is replaced by a spatially varying field after the chain's center of mass has travelled a distance equal to $d=270$ (note that this includes several passages through the periodic cell). The inhomogeneous fields then acts until the end of the simulation, during which $d$ reaches values as large as $340$.  We estimate $E_0$ using the previous simulations as average of the converging field along the centerline of the pore from the capture radius to the entrance. This yields typical values around $E_0=\bar{E}\approx 0.5$. Other parameters and conditions of these simulations are the same as the previous ones.

Simulations are performed for four different situations: 40 simulations without HI and counterions (NH-NC), 40 simulations without HI but with counterions (NH-WC), 120 simulations with HI but without counterions (WH-NC) and finally 100 simulations both with HI and counterions (WH-WC).

\begin{SCfigure}
\begin{picture} (152,240)
\put(-5,-8) {
\includegraphics[width=0.64\columnwidth]{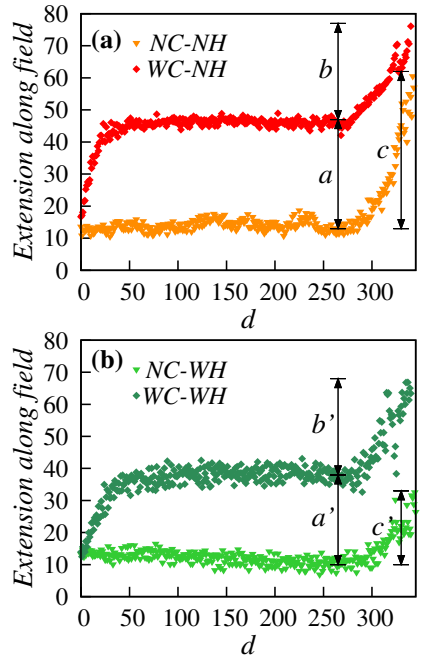}}
\put(30,225){\textbf{(a)}}
\put(30,104){\textbf{(b)}}
\end{picture}
\caption{
Extension of a polymer ($N=30$) along field lines versus the position of the polymer's center of mass. (a) In the absence of HI  (NH), polymer elongates in a homogeneous field ($d<270$) only if counterions are present (WC). In contrast, an inhomogeneous field ($d>270$) is able to deform the chain even in the absence of counterions (NC). (b) Similar data as in (a) but with HI (WH). The general behavior is essentially the same as in (a).}
\label{Extension_tot}
\end{SCfigure}

Results obtained from these simulations are shown in Fig.~\ref{Extension_tot}, where the extension of a polymer ($N=30$) along field lines is plotted versus the position of the polymer's center of mass. Interestingly, regardless of the presence or absence of hydrodynamic interactions, a uniform electric field can deform a charged polymer only if counterions are present in the system. This best seen by considering the data for $d<270$, where the applied field is homogeneous. An inhomogeneous field, on the other hand ($d>270$), can deform a charged polymer both in the presence as well as in the absence of surrounding counterions. In this context, hydrodynamic interactions seem to play a minor role as they only influence the magnitude of the effect slightly. Interestingly, the data with HI exhibit significantly larger fluctuations despite the fact that the number of simulations in this case is larger than those without HI. We attribute this effect to the HI-induced anomalies described in literature \cite{Frank2008,Manghi2006}. These anomalies also show themselves in a small elongation perpendicular to the field lines of the polymer in the absence of counterions (negative slope for $d<270$ in Fig.~\ref{Extension_tot}b).

Moreover, as the bare polyelectrolyte (no counterions) reaches the converging field, it experiences an elongation which is significantly larger in the absence of HI as compared to the case with hydrodynamic interactions (compare $c$ and $c{'}$ in Fig.~\ref{Extension_tot}). This shows that, in the absence of counterions, field gradient effects are considerably overemphasized if one does not take account of hydrodynamic interactions. On the other hand, in the more realistic case, where counterions are present in the system, the absence of HI only leads to a slight increase of the deformation behavior both in a uniform as well as non-uniform electric field.

\section{Conclusions}
In this paper, we study DNA deformation before reaching the pore in translocation phenomenon and investigate its effect on the capture process by taking into account the electrostatic and hydrodynamic interactions.
An analytical form for the electric field near the pore is obtained by approximating the pore shape with a one-sheeted hyperboloid.
An approximate expression for the electric potential, as a function of the applied voltage and the pore characteristics, is also given for the entire whole space. This provides us with sufficient information to find a theoretical description for DNA elongation.
By applying the Long's electrohydrodynamic equivalence and Randall's kinematic model for deformation, the velocity gradient tensor and hence the extension/compression axes of DNA during the capture process are determined. It is then shown that, before entering the pore, the molecule slightly stretches along the symmetry axis of the pore and, after passing through the pore, the stretched DNA experiences a compression along the same axis.

Using a coarse-grained bead-spring model and considering the hydrodynamic and electrostatic interactions, we then study via a hybrid lattice Boltzmann-molecular dynamics (LB-MD) approach, the deformation of a DNA molecule in the vicinity of the pore and discuss its relation to the dynamics of translocation.
The extension-strain curves show the effective DNA deformation before reaching the pore. It is argued that this helps the end monomers to find the pore entrance more easily, thus facilitating the translocation process.
Our simulations show that nearly always, DNA approaches the pore with one of its ends closer to the pore and hence most of DNA approaches result in successful translocations.

Furthermore, by a comparison of simulations with explicit counterions with those where counterions are replaced by an effective monomer charge, we provide evidence that such a mean-field model for the chain charge significantly underestimates the approaching probability of the end monomers. These findings are shown to be in line with experimental studies of the capture process.

One important issue addressed in this study regards the interplay between hydrodynamic interactions, presence of counterions and electric field gradients. A result of these studies is that, in the absence of counterions, a strong coupling between hydrodynamic interactions and electric field gradient occurs. In particular, field gradient effects are overemphasized if both counterions and HI are absent. On the other hand, if one properly takes account of counterions, hydrodynamic interactions lead to a more accurate description of the system behavior but do not significantly change the overall picture.

Finally, it is worth mentioning that our proposed electric field makes a connection between two distinct fields of research in nanopore translocation: (1) Studies in which the access resistance of a nanopore is investigated and the electric potential is assumed to drop mainly inside the pore \cite{Hall1975,Kowalczyk2011} and (2) Those in which the effect of the electric field outside the pore on the polyelectrolyte diffusion is studied and nearly always the spatial continuity of the electric field or potential at the pore mouth is neglected \cite{Hatlo2010,Grosberg2010}.

\section{Acknowledgement}
The authors are gratefully thankful to Dr Timm Kr\"uger and Markus Gross for their useful discussions about the hybrid simulation of MD-LB method. We also thank Stefan Kesselheim
for his helpful remarks and N. Heydari for editing the manuscript of this article. F. F. acknowledges the kind hospitality of the ICAMS during her visits. We also thank the anonymous referee whose critical comments helped us in improving the quality of our manuscript.
The simulations presented in this paper were carried out on the High Performance Computing 
Cluster supported by the computer science department of the Institute for Research in Fundamental Sciences (IPM).
\footnotesize{
\bibliography{biblio} 

\providecommand*{\mcitethebibliography}{\thebibliography}
\csname @ifundefined\endcsname{endmcitethebibliography}
{\let\endmcitethebibliography\endthebibliography}{}
\begin{mcitethebibliography}{56}
\providecommand*{\natexlab}[1]{#1}
\providecommand*{\mciteSetBstSublistMode}[1]{}
\providecommand*{\mciteSetBstMaxWidthForm}[2]{}
\providecommand*{\mciteBstWouldAddEndPuncttrue}
  {\def\EndOfBibitem{\unskip.}}
\providecommand*{\mciteBstWouldAddEndPunctfalse}
  {\let\EndOfBibitem\relax}
\providecommand*{\mciteSetBstMidEndSepPunct}[3]{}
\providecommand*{\mciteSetBstSublistLabelBeginEnd}[3]{}
\providecommand*{\EndOfBibitem}{}
\mciteSetBstSublistMode{f}
\mciteSetBstMaxWidthForm{subitem}
{(\emph{\alph{mcitesubitemcount}})}
\mciteSetBstSublistLabelBeginEnd{\mcitemaxwidthsubitemform\space}
{\relax}{\relax}

\bibitem[Meller(2003)]{Meller2003}
A.~Meller, \emph{Journal of Physics: Condensed Matter}, 2003, \textbf{15},
  R581--R607\relax
\mciteBstWouldAddEndPuncttrue
\mciteSetBstMidEndSepPunct{\mcitedefaultmidpunct}
{\mcitedefaultendpunct}{\mcitedefaultseppunct}\relax
\EndOfBibitem
\bibitem[Nakane \emph{et~al.}(2002)Nakane, Akeson, and Marziali]{Nakane2002}
J.~Nakane, M.~Akeson and A.~Marziali, \emph{Electrophoresis}, 2002,
  \textbf{23}, 2592--2601\relax
\mciteBstWouldAddEndPuncttrue
\mciteSetBstMidEndSepPunct{\mcitedefaultmidpunct}
{\mcitedefaultendpunct}{\mcitedefaultseppunct}\relax
\EndOfBibitem
\bibitem[Meller and Branton(2002)]{Meller2002}
A.~Meller and D.~Branton, \emph{Electrophoresis}, 2002, \textbf{23},
  2583--2591\relax
\mciteBstWouldAddEndPuncttrue
\mciteSetBstMidEndSepPunct{\mcitedefaultmidpunct}
{\mcitedefaultendpunct}{\mcitedefaultseppunct}\relax
\EndOfBibitem
\bibitem[Meller \emph{et~al.}(2001)Meller, Nivon, and Branton]{Meller2001}
A.~Meller, L.~Nivon and D.~Branton, \emph{Physical Review Letters}, 2001,
  \textbf{86}, 3435--3438\relax
\mciteBstWouldAddEndPuncttrue
\mciteSetBstMidEndSepPunct{\mcitedefaultmidpunct}
{\mcitedefaultendpunct}{\mcitedefaultseppunct}\relax
\EndOfBibitem
\bibitem[Chang \emph{et~al.}(2004)Chang, Kosari, Andreadakis, Alam, Vasmatzis,
  and Bashir]{Chang2004}
H.~Chang, F.~Kosari, G.~Andreadakis, M.~A. Alam, G.~Vasmatzis and R.~Bashir,
  \emph{Nano Letters}, 2004, \textbf{4}, 1551--1556\relax
\mciteBstWouldAddEndPuncttrue
\mciteSetBstMidEndSepPunct{\mcitedefaultmidpunct}
{\mcitedefaultendpunct}{\mcitedefaultseppunct}\relax
\EndOfBibitem
\bibitem[Storm \emph{et~al.}(2005)Storm, Storm, Chen, Zandbergen, Joanny, and
  Dekker]{Storm2005}
A.~J. Storm, C.~Storm, J.~Chen, H.~Zandbergen, J.-F. Joanny and C.~Dekker,
  \emph{Nano Letters}, 2005, \textbf{5}, 1193--1197\relax
\mciteBstWouldAddEndPuncttrue
\mciteSetBstMidEndSepPunct{\mcitedefaultmidpunct}
{\mcitedefaultendpunct}{\mcitedefaultseppunct}\relax
\EndOfBibitem
\bibitem[Bezrukov and Vodyanoy(1992)]{Bezrukov1992}
S.~M. Bezrukov and I.~Vodyanoy, \emph{Biophysical journal}, 1992, \textbf{62},
  10--11\relax
\mciteBstWouldAddEndPuncttrue
\mciteSetBstMidEndSepPunct{\mcitedefaultmidpunct}
{\mcitedefaultendpunct}{\mcitedefaultseppunct}\relax
\EndOfBibitem
\bibitem[Ambjörnsson \emph{et~al.}(2002)Ambjörnsson, Apell, Konkoli, {Di
  Marzio}, and Kasianowicz]{Ambjornsson2002}
T.~Ambjörnsson, S.~P. Apell, Z.~Konkoli, E.~A. {Di Marzio} and J.~J.
  Kasianowicz, \emph{The Journal of Chemical Physics}, 2002, \textbf{117},
  4063\relax
\mciteBstWouldAddEndPuncttrue
\mciteSetBstMidEndSepPunct{\mcitedefaultmidpunct}
{\mcitedefaultendpunct}{\mcitedefaultseppunct}\relax
\EndOfBibitem
\bibitem[Loebl \emph{et~al.}(2003)Loebl, Randel, Goodwin, and
  Matthai]{Loebl2003}
H.~C. Loebl, R.~Randel, S.~P. Goodwin and C.~C. Matthai, \emph{Physical Review
  E}, 2003, \textbf{67}, 041913\relax
\mciteBstWouldAddEndPuncttrue
\mciteSetBstMidEndSepPunct{\mcitedefaultmidpunct}
{\mcitedefaultendpunct}{\mcitedefaultseppunct}\relax
\EndOfBibitem
\bibitem[Chern \emph{et~al.}(2001)Chern, Cárdenas, and Coalson]{Chern2001}
S.~S. Chern, A.~E. Cárdenas and R.~D. Coalson, \emph{The Journal of Chemical
  Physics}, 2001, \textbf{115}, 7772\relax
\mciteBstWouldAddEndPuncttrue
\mciteSetBstMidEndSepPunct{\mcitedefaultmidpunct}
{\mcitedefaultendpunct}{\mcitedefaultseppunct}\relax
\EndOfBibitem
\bibitem[Wanunu \emph{et~al.}(2010)Wanunu, Morrison, Rabin, Grosberg, and
  Meller]{Wanunu2010}
M.~Wanunu, W.~Morrison, Y.~Rabin, A.~Y. Grosberg and A.~Meller, \emph{Nature
  Nanotechnology}, 2010, \textbf{5}, 160--165\relax
\mciteBstWouldAddEndPuncttrue
\mciteSetBstMidEndSepPunct{\mcitedefaultmidpunct}
{\mcitedefaultendpunct}{\mcitedefaultseppunct}\relax
\EndOfBibitem
\bibitem[Grosberg and Rabin(2010)]{Grosberg2010}
A.~Y. Grosberg and Y.~Rabin, \emph{The Journal of chemical physics}, 2010,
  \textbf{133}, 165102\relax
\mciteBstWouldAddEndPuncttrue
\mciteSetBstMidEndSepPunct{\mcitedefaultmidpunct}
{\mcitedefaultendpunct}{\mcitedefaultseppunct}\relax
\EndOfBibitem
\bibitem[Muthukumar(2010)]{Muthukumar2010}
M.~Muthukumar, \emph{The Journal of chemical physics}, 2010, \textbf{132},
  195101\relax
\mciteBstWouldAddEndPuncttrue
\mciteSetBstMidEndSepPunct{\mcitedefaultmidpunct}
{\mcitedefaultendpunct}{\mcitedefaultseppunct}\relax
\EndOfBibitem
\bibitem[Chen \emph{et~al.}(2004)Chen, Gu, Brandin, Kim, Wang, and
  Branton]{Chen2004}
P.~Chen, J.~Gu, E.~Brandin, Y.-R. Kim, Q.~Wang and D.~Branton, \emph{Nano
  Letters}, 2004, \textbf{4}, 2293--2298\relax
\mciteBstWouldAddEndPuncttrue
\mciteSetBstMidEndSepPunct{\mcitedefaultmidpunct}
{\mcitedefaultendpunct}{\mcitedefaultseppunct}\relax
\EndOfBibitem
\bibitem[Wong and Muthukumar(2007)]{Wong2007}
C.~T.~A. Wong and M.~Muthukumar, \emph{The Journal of chemical physics}, 2007,
  \textbf{126}, 164903\relax
\mciteBstWouldAddEndPuncttrue
\mciteSetBstMidEndSepPunct{\mcitedefaultmidpunct}
{\mcitedefaultendpunct}{\mcitedefaultseppunct}\relax
\EndOfBibitem
\bibitem[Perkins \emph{et~al.}(1995)Perkins, Smith, Larson, and
  Chu]{Perkins1995}
T.~T. Perkins, D.~E. Smith, R.~G. Larson and S.~Chu, \emph{Science}, 1995,
  \textbf{268}, 83--87\relax
\mciteBstWouldAddEndPuncttrue
\mciteSetBstMidEndSepPunct{\mcitedefaultmidpunct}
{\mcitedefaultendpunct}{\mcitedefaultseppunct}\relax
\EndOfBibitem
\bibitem[Smith \emph{et~al.}(1996)Smith, Cui, and Bustamante]{Smith1996}
S.~B. Smith, Y.~Cui and C.~Bustamante, \emph{Science}, 1996, \textbf{271},
  795--799\relax
\mciteBstWouldAddEndPuncttrue
\mciteSetBstMidEndSepPunct{\mcitedefaultmidpunct}
{\mcitedefaultendpunct}{\mcitedefaultseppunct}\relax
\EndOfBibitem
\bibitem[Bakajin \emph{et~al.}(1998)Bakajin, Duke, Chou, Chan, Austin, and
  Cox]{Bakajin1998}
O.~B. Bakajin, T.~A.~J. Duke, C.~F. Chou, S.~S. Chan, R.~H. Austin and E.~C.
  Cox, \emph{Physical Review Letters}, 1998, \textbf{80}, 2737--2740\relax
\mciteBstWouldAddEndPuncttrue
\mciteSetBstMidEndSepPunct{\mcitedefaultmidpunct}
{\mcitedefaultendpunct}{\mcitedefaultseppunct}\relax
\EndOfBibitem
\bibitem[Han \emph{et~al.}(1999)Han, Turner, and Craighead]{Han1999}
J.~Han, S.~W. Turner and H.~G. Craighead, \emph{Physical Review Letters}, 1999,
  \textbf{83}, 1688 --1691\relax
\mciteBstWouldAddEndPuncttrue
\mciteSetBstMidEndSepPunct{\mcitedefaultmidpunct}
{\mcitedefaultendpunct}{\mcitedefaultseppunct}\relax
\EndOfBibitem
\bibitem[Bertrand and Slater(2007)]{Bertrand2007}
M.~Bertrand and G.~W. Slater, \emph{The European physical journal E}, 2007,
  \textbf{23}, 83--89\relax
\mciteBstWouldAddEndPuncttrue
\mciteSetBstMidEndSepPunct{\mcitedefaultmidpunct}
{\mcitedefaultendpunct}{\mcitedefaultseppunct}\relax
\EndOfBibitem
\bibitem[Brochard-Wyart(1993)]{Brochard-Wyart1993}
F.~Brochard-Wyart, \emph{Europhysics Letters}, 1993, \textbf{23},
  105--111\relax
\mciteBstWouldAddEndPuncttrue
\mciteSetBstMidEndSepPunct{\mcitedefaultmidpunct}
{\mcitedefaultendpunct}{\mcitedefaultseppunct}\relax
\EndOfBibitem
\bibitem[Ferree and Blanch(2003)]{Ferree2003}
S.~Ferree and H.~W. Blanch, \emph{Biophysical Journal}, 2003, \textbf{85},
  2539--2546\relax
\mciteBstWouldAddEndPuncttrue
\mciteSetBstMidEndSepPunct{\mcitedefaultmidpunct}
{\mcitedefaultendpunct}{\mcitedefaultseppunct}\relax
\EndOfBibitem
\bibitem[Larson \emph{et~al.}(2006)Larson, Yantz, Zhong, Charnas, Antoni,
  Gallo, Gillis, Neely, Phillips, Wong, Gullans, and Gilmanshin]{Larson2006}
J.~W. Larson, G.~R. Yantz, Q.~Zhong, R.~Charnas, C.~M.~D. Antoni, M.~V. Gallo,
  K.~A. Gillis, L.~A. Neely, K.~M. Phillips, G.~G. Wong, S.~R. Gullans and
  R.~Gilmanshin, \emph{Lab on a Chip}, 2006, \textbf{6}, 1187--1199\relax
\mciteBstWouldAddEndPuncttrue
\mciteSetBstMidEndSepPunct{\mcitedefaultmidpunct}
{\mcitedefaultendpunct}{\mcitedefaultseppunct}\relax
\EndOfBibitem
\bibitem[de~Gennes(1979)]{deGennes1999}
P.~G. de~Gennes, \emph{{S}caling {C}oncepts in {P}olymer {P}hysics}, Cornell
  University, New York, 1979\relax
\mciteBstWouldAddEndPuncttrue
\mciteSetBstMidEndSepPunct{\mcitedefaultmidpunct}
{\mcitedefaultendpunct}{\mcitedefaultseppunct}\relax
\EndOfBibitem
\bibitem[Doi and Edwards(1986)]{DoiEdwards1986}
M.~Doi and S.~F. Edwards, \emph{{T}he {T}heory of {P}olymer {D}ynamics}, Oxford
  University, Oxford, 1986\relax
\mciteBstWouldAddEndPuncttrue
\mciteSetBstMidEndSepPunct{\mcitedefaultmidpunct}
{\mcitedefaultendpunct}{\mcitedefaultseppunct}\relax
\EndOfBibitem
\bibitem[Randall and Doyle(2005)]{Randall2005}
G.~C. Randall and P.~S. Doyle, \emph{Macromolecules}, 2005, \textbf{38},
  2410--2418\relax
\mciteBstWouldAddEndPuncttrue
\mciteSetBstMidEndSepPunct{\mcitedefaultmidpunct}
{\mcitedefaultendpunct}{\mcitedefaultseppunct}\relax
\EndOfBibitem
\bibitem[Kim and Doyle(2006)]{Kim2006}
J.~M. Kim and P.~S. Doyle, \emph{The Journal of chemical physics}, 2006,
  \textbf{125}, 074906\relax
\mciteBstWouldAddEndPuncttrue
\mciteSetBstMidEndSepPunct{\mcitedefaultmidpunct}
{\mcitedefaultendpunct}{\mcitedefaultseppunct}\relax
\EndOfBibitem
\bibitem[Tessier \emph{et~al.}(2002)Tessier, Labrie, and Slater]{Tessier2002}
F.~Tessier, J.~Labrie and G.~W. Slater, \emph{Macromolecules}, 2002,
  \textbf{35}, 4791--4800\relax
\mciteBstWouldAddEndPuncttrue
\mciteSetBstMidEndSepPunct{\mcitedefaultmidpunct}
{\mcitedefaultendpunct}{\mcitedefaultseppunct}\relax
\EndOfBibitem
\bibitem[Long \emph{et~al.}(1998)Long, Dobrynin, Rubinstein, and
  Ajdari]{Long1998}
D.~Long, A.~V. Dobrynin, M.~Rubinstein and A.~Ajdari, \emph{The Journal of
  Chemical Physics}, 1998, \textbf{108}, 1234--1244\relax
\mciteBstWouldAddEndPuncttrue
\mciteSetBstMidEndSepPunct{\mcitedefaultmidpunct}
{\mcitedefaultendpunct}{\mcitedefaultseppunct}\relax
\EndOfBibitem
\bibitem[Kowalczyk \emph{et~al.}(2011)Kowalczyk, Grosberg, Rabin, and
  Dekker]{Kowalczyk2011}
S.~W. Kowalczyk, A.~Y. Grosberg, Y.~Rabin and C.~Dekker, \emph{Nanotechnology},
  2011, \textbf{22}, 315101\relax
\mciteBstWouldAddEndPuncttrue
\mciteSetBstMidEndSepPunct{\mcitedefaultmidpunct}
{\mcitedefaultendpunct}{\mcitedefaultseppunct}\relax
\EndOfBibitem
\bibitem[Storm \emph{et~al.}(2005)Storm, Chen, Zandbergen, and
  Dekker]{Storm2005-2}
A.~J. Storm, J.~Chen, H.~Zandbergen and C.~Dekker, \emph{Physical Review E},
  2005, \textbf{71}, 051903\relax
\mciteBstWouldAddEndPuncttrue
\mciteSetBstMidEndSepPunct{\mcitedefaultmidpunct}
{\mcitedefaultendpunct}{\mcitedefaultseppunct}\relax
\EndOfBibitem
\bibitem[Mihovilovic \emph{et~al.}(2012)Mihovilovic, Hagerty, and
  Stein]{Mihovilovic2012}
M.~Mihovilovic, N.~Hagerty and D.~Stein, \emph{arXiv:1209.3250v1
  [physics.bio-ph]}, 2012\relax
\mciteBstWouldAddEndPuncttrue
\mciteSetBstMidEndSepPunct{\mcitedefaultmidpunct}
{\mcitedefaultendpunct}{\mcitedefaultseppunct}\relax
\EndOfBibitem
\bibitem[Kesselheim \emph{et~al.}(2012)Kesselheim, Sega, and
  Holm]{Kesselheim2012}
S.~Kesselheim, M.~Sega and C.~Holm, \emph{Soft Matter}, 2012, \textbf{8},
  9480--9486\relax
\mciteBstWouldAddEndPuncttrue
\mciteSetBstMidEndSepPunct{\mcitedefaultmidpunct}
{\mcitedefaultendpunct}{\mcitedefaultseppunct}\relax
\EndOfBibitem
\bibitem[Hall(1975)]{Hall1975}
J.~E. Hall, \emph{The Journal of general physiology}, 1975, \textbf{66},
  531--532\relax
\mciteBstWouldAddEndPuncttrue
\mciteSetBstMidEndSepPunct{\mcitedefaultmidpunct}
{\mcitedefaultendpunct}{\mcitedefaultseppunct}\relax
\EndOfBibitem
\bibitem[Deen(1998)]{Deen1998}
W.~M. Deen, \emph{{A}nalysis of {T}ransport {P}henomena}, Oxford University,
  New York, 1998\relax
\mciteBstWouldAddEndPuncttrue
\mciteSetBstMidEndSepPunct{\mcitedefaultmidpunct}
{\mcitedefaultendpunct}{\mcitedefaultseppunct}\relax
\EndOfBibitem
\bibitem[Limbach \emph{et~al.}(2006)Limbach, Arnold, Mann, and
  Holm]{Limbach2006}
H.~Limbach, A.~Arnold, B.~Mann and C.~Holm, \emph{Computer Physics
  Communications}, 2006, \textbf{174}, 704--727\relax
\mciteBstWouldAddEndPuncttrue
\mciteSetBstMidEndSepPunct{\mcitedefaultmidpunct}
{\mcitedefaultendpunct}{\mcitedefaultseppunct}\relax
\EndOfBibitem
\bibitem[McNamara and Zanetti(1988)]{McNamara1988}
G.~R. McNamara and G.~Zanetti, \emph{Physical Review Letters}, 1988,
  \textbf{61}, 2332--2335\relax
\mciteBstWouldAddEndPuncttrue
\mciteSetBstMidEndSepPunct{\mcitedefaultmidpunct}
{\mcitedefaultendpunct}{\mcitedefaultseppunct}\relax
\EndOfBibitem
\bibitem[Higuera and Jimenez(1989)]{Higuera1989}
F.~Higuera and J.~Jimenez, \emph{Europhysics Letters}, 1989, \textbf{9},
  663--668\relax
\mciteBstWouldAddEndPuncttrue
\mciteSetBstMidEndSepPunct{\mcitedefaultmidpunct}
{\mcitedefaultendpunct}{\mcitedefaultseppunct}\relax
\EndOfBibitem
\bibitem[Benzi \emph{et~al.}(1992)Benzi, Succi, and Vergassola]{Benzi1992}
R.~Benzi, S.~Succi and M.~Vergassola, \emph{Physics Reports}, 1992,
  \textbf{222}, 145--197\relax
\mciteBstWouldAddEndPuncttrue
\mciteSetBstMidEndSepPunct{\mcitedefaultmidpunct}
{\mcitedefaultendpunct}{\mcitedefaultseppunct}\relax
\EndOfBibitem
\bibitem[Qian \emph{et~al.}(1992)Qian, D'Humieres, and Lallemand]{Qian1992}
Y.~Qian, D.~D'Humieres and P.~Lallemand, \emph{Europhysics Letters}, 1992,
  \textbf{17}, 479\relax
\mciteBstWouldAddEndPuncttrue
\mciteSetBstMidEndSepPunct{\mcitedefaultmidpunct}
{\mcitedefaultendpunct}{\mcitedefaultseppunct}\relax
\EndOfBibitem
\bibitem[Gross \emph{et~al.}(2010)Gross, Adhikari, Cates, and
  Varnik]{Gross2010}
M.~Gross, R.~Adhikari, M.~E. Cates and F.~Varnik, \emph{Physical Review E},
  2010, \textbf{82}, 056714\relax
\mciteBstWouldAddEndPuncttrue
\mciteSetBstMidEndSepPunct{\mcitedefaultmidpunct}
{\mcitedefaultendpunct}{\mcitedefaultseppunct}\relax
\EndOfBibitem
\bibitem[Gross \emph{et~al.}(2011)Gross, Cates, Varnik, and
  Adhikari]{Gross2011}
M.~Gross, M.~E. Cates, F.~Varnik and R.~Adhikari, \emph{Journal of Statistical
  Mechanics}, 2011, \textbf{03}, P03030\relax
\mciteBstWouldAddEndPuncttrue
\mciteSetBstMidEndSepPunct{\mcitedefaultmidpunct}
{\mcitedefaultendpunct}{\mcitedefaultseppunct}\relax
\EndOfBibitem
\bibitem[Deserno and Holm(1998)]{Deserno1998}
M.~Deserno and C.~Holm, \emph{The Journal of Chemical Physics}, 1998,
  \textbf{109}, 7694--7701\relax
\mciteBstWouldAddEndPuncttrue
\mciteSetBstMidEndSepPunct{\mcitedefaultmidpunct}
{\mcitedefaultendpunct}{\mcitedefaultseppunct}\relax
\EndOfBibitem
\bibitem[Arnold \emph{et~al.}(2006)Arnold, Mann, and Holm]{Arnold2006}
A.~Arnold, B.~A.~F. Mann and C.~Holm, \emph{Lecture Notes in Physics}, 2006,
  \textbf{703}, 193--221\relax
\mciteBstWouldAddEndPuncttrue
\mciteSetBstMidEndSepPunct{\mcitedefaultmidpunct}
{\mcitedefaultendpunct}{\mcitedefaultseppunct}\relax
\EndOfBibitem
\bibitem[Lansac \emph{et~al.}(2004)Lansac, Maiti, and Glaser]{Lansac2004}
Y.~Lansac, P.~K. Maiti and M.~A. Glaser, \emph{Polymer}, 2004, \textbf{45},
  3099--3110\relax
\mciteBstWouldAddEndPuncttrue
\mciteSetBstMidEndSepPunct{\mcitedefaultmidpunct}
{\mcitedefaultendpunct}{\mcitedefaultseppunct}\relax
\EndOfBibitem
\bibitem[Grass and Holm(2008)]{Grass2008}
K.~Grass and C.~Holm, \emph{Journal of Physics: Condensed Matter}, 2008,
  \textbf{20}, 494217\relax
\mciteBstWouldAddEndPuncttrue
\mciteSetBstMidEndSepPunct{\mcitedefaultmidpunct}
{\mcitedefaultendpunct}{\mcitedefaultseppunct}\relax
\EndOfBibitem
\bibitem[Ahlrichs and Dunweg(1999)]{Ahlrichs1999}
P.~Ahlrichs and B.~Dunweg, \emph{The Journal of chemical physics}, 1999,
  \textbf{111}, 8225--8239\relax
\mciteBstWouldAddEndPuncttrue
\mciteSetBstMidEndSepPunct{\mcitedefaultmidpunct}
{\mcitedefaultendpunct}{\mcitedefaultseppunct}\relax
\EndOfBibitem
\bibitem[Ladd \emph{et~al.}(2009)Ladd, Kekre, and Butler]{Ladd2009}
A.~J.~C. Ladd, R.~Kekre and J.~E. Butler, \emph{Physical Review E}, 2009,
  \textbf{80}, 036704\relax
\mciteBstWouldAddEndPuncttrue
\mciteSetBstMidEndSepPunct{\mcitedefaultmidpunct}
{\mcitedefaultendpunct}{\mcitedefaultseppunct}\relax
\EndOfBibitem
\bibitem[Pham \emph{et~al.}(2009)Pham, Schiller, Prakash, and
  Duenweg]{Pham2009}
T.~T. Pham, U.~D. Schiller, J.~R. Prakash and B.~Duenweg, \emph{The Journal of
  chemical physics}, 2009, \textbf{131}, 164114\relax
\mciteBstWouldAddEndPuncttrue
\mciteSetBstMidEndSepPunct{\mcitedefaultmidpunct}
{\mcitedefaultendpunct}{\mcitedefaultseppunct}\relax
\EndOfBibitem
\bibitem[Varnik(2000)]{Varnik2000b}
F.~Varnik, \emph{PhD thesis}, University of Mainz, 2000\relax
\mciteBstWouldAddEndPuncttrue
\mciteSetBstMidEndSepPunct{\mcitedefaultmidpunct}
{\mcitedefaultendpunct}{\mcitedefaultseppunct}\relax
\EndOfBibitem
\bibitem[Witten and Schafer(1981)]{Witten1981}
T.~A. Witten and L.~Schafer, \emph{The Journal of Chemical Physics}, 1981,
  \textbf{74}, 2582--2588\relax
\mciteBstWouldAddEndPuncttrue
\mciteSetBstMidEndSepPunct{\mcitedefaultmidpunct}
{\mcitedefaultendpunct}{\mcitedefaultseppunct}\relax
\EndOfBibitem
\bibitem[Wu \emph{et~al.}(2011)Wu, Wei, and Hsiao]{Wu2011}
K.-M. Wu, Y.-F. Wei and P.-Y. Hsiao, \emph{Electrophoresis}, 2011, \textbf{32},
  3348--3363\relax
\mciteBstWouldAddEndPuncttrue
\mciteSetBstMidEndSepPunct{\mcitedefaultmidpunct}
{\mcitedefaultendpunct}{\mcitedefaultseppunct}\relax
\EndOfBibitem
\bibitem[Netz(2003)]{Netz2003}
R.~R. Netz, \emph{Journal of Physical Chemistry B}, 2003, \textbf{107},
  8208--8217\relax
\mciteBstWouldAddEndPuncttrue
\mciteSetBstMidEndSepPunct{\mcitedefaultmidpunct}
{\mcitedefaultendpunct}{\mcitedefaultseppunct}\relax
\EndOfBibitem
\bibitem[Frank and Winkler(2008)]{Frank2008}
S.~Frank and R.~G. Winkler, \emph{EPL (Europhysics Letters)}, 2008,
  \textbf{83}, 38004\relax
\mciteBstWouldAddEndPuncttrue
\mciteSetBstMidEndSepPunct{\mcitedefaultmidpunct}
{\mcitedefaultendpunct}{\mcitedefaultseppunct}\relax
\EndOfBibitem
\bibitem[Manghi \emph{et~al.}(2006)Manghi, Schlagberger, Kim, and
  Netz]{Manghi2006}
M.~Manghi, X.~Schlagberger, Y.-W. Kim and R.~R. Netz, \emph{Soft Matter}, 2006,
  \textbf{2}, 653--668\relax
\mciteBstWouldAddEndPuncttrue
\mciteSetBstMidEndSepPunct{\mcitedefaultmidpunct}
{\mcitedefaultendpunct}{\mcitedefaultseppunct}\relax
\EndOfBibitem
\bibitem[Hatlo \emph{et~al.}(2011)Hatlo, Panja, and {Van Roij}]{Hatlo2010}
M.~M. Hatlo, D.~Panja and R.~{Van Roij}, \emph{Physical Review Letters}, 2011,
  \textbf{107}, 068101\relax
\mciteBstWouldAddEndPuncttrue
\mciteSetBstMidEndSepPunct{\mcitedefaultmidpunct}
{\mcitedefaultendpunct}{\mcitedefaultseppunct}\relax
\EndOfBibitem
\end{mcitethebibliography}
\bibliographystyle{rsc} 
}

\end{document}